\newcommand{\be}{\begin{equation}}
\newcommand{\ee}{\end{equation}}
\newcommand{\bea}{\begin{eqnarray}}
\newcommand{\eea}{\end{eqnarray}}
\documentclass[aps,pra,amsmath,amssymb,amsfonts,onecolumn,nofootinbib]{revtex4}
\usepackage{graphicx}% Include figure files
\usepackage{dcolumn}% Align table columns on decimal point
\usepackage{bm}% bold math
\usepackage{amsmath}
\usepackage{amssymb}
\usepackage{float}
\def\d{d\kern-.8 ex\vrule height 1.3 ex depth-1.24 ex width .7 ex \kern .15 ex}
\def\D{D\kern-1.7 ex\vrule height .87 ex depth-.8 ex width .7 ex \kern .95 ex}

\begin{document}
\title{Spontaneous isotropy breaking for vortices in nonlinear left-handed metamaterials}

\author{Trivko Kukolj}
\email{trivko98@gmail.com}
\affiliation{Scientific Computing Lab, Center for the Study of Complex Systems, Institute of Physics Belgrade, University of Belgrade, Serbia}
\affiliation{Department of Physics, Faculty of Sciences, University of Novi Sad, Trg Dositeja Obradovi\'ca 4, Novi Sad, Serbia}

\author{Mihailo~\v{C}ubrovi\'{c}}
\email{mcubrovic@gmail.com}
\affiliation{Scientific Computing Lab, Center for the Study of Complex Systems, Institute of Physics Belgrade, University of Belgrade, Serbia}

\date{\today}

\begin{abstract}
We explore numerically and analytically the pattern formation and symmetry breaking of beams propagating through left-handed (negative) nonlinear metamaterials. When the input beam is a vortex with topological charge (winding number) $Q$, the initially circular (isotropic) beam acquires the symmetry of a polygon with $Q$, $2Q$ or $3Q$ sides, depending on the details of the response functions of the material. Within an effective field-theory model, this phenomenon turns out to be a case of spontaneous dynamical symmetry breaking described by a Landau-Ginzburg functional. Complex nonlinear dependence of the magnetic permittivity on the magnetic field of the beam plays a central role, as it introduces branch cuts in the mean-field solution, and permutations among different branches give rise to discrete symmetries of the patterns. By considering loop corrections in the effective Landau-Ginzburg field theory we obtain reasonably accurate predictions of the numerical results.
\end{abstract}

%\pacs{05.70.Fh,42.65.Sf,64.60.De,64.70.qd,75.10.Nr}

\maketitle

\section{Introduction}

% Nefative metamat - intro
The idea of a material with negative refraction index was first considered long before it could be realized in experiment, in the now famous paper by Veselago \cite{veselago}, in 1968. He considered a material with negative electric permeability $\epsilon$ \emph{and} magnetic permittivity $\mu$, and predicted a number of interesting properties in such systems, among them negative refraction. Only much later did it become possible to combine elements with negative $\epsilon$ and negative $\mu$ at a microscopic level, as a composite metamaterial. First experimental realizations were reported in \cite{first,exper}. Negativity, or left-handedness, is typically only achieved in a narrow frequency range, close to the resonant frequency of the conductive elements of the metamaterial. This was the original motivation for studying nonlinear effects in these systems. Nonlinearities can be strengthened by appropriate design at the microscopic level. The study of nonlinear phenomena in metamaterials started with \cite{zarov03}. This has become a broad and important field in metamaterials research \cite{rmp}. Nonlinear phenomena like solitons \cite{zarova05,sadriv05}, nonlinear surface waves \cite{sadriv04}, modulational instability \cite{wen,walasik} and ultrashort pulses \cite{tsitsas} were observed. Other work in left-handed metamaterials relevant for our paper includes, among others, \cite{sadriv03,kivstamm,ma,fan,korotki,sarma,kevrek,ciral,symmbreak}. We have no intention of being exhaustive in this short review of the literature; we merely mention the results we have directly used or found particularly inspiring.

% Anizotropija - nasa poenta
The focus of our work is the dynamics of symmetry breaking in intensity patterns of electromagnetic waves propagating through a left-handed nonlinear metamaterial. Numerical solutions of the equations of motion reveal that circular (usually Gaussian) input beams turn into polygonal patterns, with some discrete symmetry. This fits the textbook notion of symmetry breaking, more specifically dynamical symmetry breaking. The general theory of dynamical criticality is by now well-developed \cite{cross} and has been applied to numerous systems \cite{rabin}. In \cite{rabin}, a systematic theory of isotropy-breaking transition is presented, though mainly for periodic and quasiperiodic structures (convection in fluids, fluctuations in quasicrystals). The basic mechanism is that the system develops momentum eigenmodes of fixed module but with multiple discrete directions on the sphere $\vert\mathbf{k}\vert=\mathrm{const.}$ in momentum space. In nonlinear negative materials, the situation is complicated by the strong frequency dependence of the magnetic permittivity but the same basic logic remains. At a fundamental level, this situation can be understood from the viewpoint of a spatially non-uniform Landau-Ginzburg theory. Quantitative accuracy is however hard to achieve; this requires cumbersome perturbative calculations. Ultimately, numerical work is the best way to describe the patterns in detail; they look like polygons or, occasionally, necklaces, with $C_{3Q}$, $C_{2Q}$ or $C_Q$ symmetry, depending on the parameter regime; here, $Q$ is the topological charge of the beam, a property we will discuss in detail in the next paragraph. The paper \cite{walasik} is very important in this context: it starts from the model derived in \cite{wen} and studies mainly necklace configurations, which consist of discrete beads (spots of high intensity) distributed more or less uniformly along a circle. The authors find the same $C_{3Q}$ symmetry that we see. Our goal is to gain a detailed understanding of the phenomenon, and move 
beyond single beams toward collective behavior and interactions.

%Vortices - intro
We have chosen to study this phenomenon on vortices, topologically nontrivial solutions where the phase of the complex electric and magnetic field winds one or more times along a closed line encircling the vortex core. Vortices appear in many systems described by a complex field, i.e., a field with $U(1)$ phase invariance \cite{vortbook,kleinert}. In optics, this is just the complex beam envelope of the electric and magnetic field. The phase of any complex wavefunction or field can wind along some closed line around a defect, forming a vortex. Famously, vortices may coexist with the superconducting order ($U(1)$ symmetry breaking) in type-II superconductors or they may exist only in the normal phase, upon destroying the superconductivity (type I). Pattern-forming systems like fluids and soft matter often have rich vortex dynamics \cite{rabin}. Other examples of vortex matter in nature arise in liquid helium \cite{liqhel}, Bose-Einstein condensates \cite{bec} and magnetic systems \cite{vortcupsach}. In two spatial dimensions, interactions among the vortices lead to a vortex unbinding phase transition of infinite order found by Berezinsky, Kosterlitz and Thouless for the planar XY model \cite{kosterthoul}. We study a three-dimensional metamaterial but with an elongated geometry, so we treat it as a $2+1$-dimensional system (the $x$ and $y$ coordinates are spatial dimensions and the $z$-direction has the formal role of time). We therefore have a similar situation to the XY model but with different equations of motion and different phenomena.

% Jos o nasem sistemu
In addition to direct numerical and analytical study of the equations of motion, we also propose an effective field-theory Lagrangian which gives slightly different equations but captures the key properties of the system. The Lagrangian form makes it easier to understand some of the phenomenology we find in numerical simulations; the foundations of the symmetry breaking are obtained from this model in a natural way. Numerical work is done with original equations of motion, as they are directly grounded in the microscopic physics. The Lagrangian is just a phenomenological tool to facilitate the theoretical understanding. It is difficult (and perhaps impossible) to package the exact
original equations in a Lagrangian form because the system is strongly nonlinear \emph{and} dissipative. Dissipative systems can be encapsulated in a Lagrangian (our Lagrangian is also dissipative) but with some limitations, and there is certainly no general method to write down a Lagrangian for a broad class of dissipative systems.

%Plan rada
The structure of the paper is the following. In the next section we describe the model of a nonlinear left-handed metamaterial, following closely the wave propagation equations used in previous research, e.g. in \cite{zarov03,zarova05,sadriv05} and others, which correspond to a specific experimentally realizable metamaterial. We also formulate and motivate the field-theory model of the system. In the third section, we describe our numerical findings, above all the anisotropy of the intensity patterns. The fourth section offers the theoretical explanation for the patterns: first by a direct approximate solution of the propagation equations, and then also from field theory, which makes the physics of the symmetry breaking particularly clear. In the fifth section we briefly discuss how to check our predictions in experiment and how prominent the effects of symmetry breaking are compared to other possible instabilities in realistic metamaterials. The last section sums up the conclusions. We have included some long calculations in the Appendices.

\section{Wave equations in a nonlinear left-handed metamaterial}

We adopt the model of \cite{zarov03,sadriv05} to describe a left-handed metamaterial with a nonlinear response. Microscopically, the material is realized as a lattice of split-ring resonators and wires. In the terahertz range, this is an experimentally well-studied system \cite{exper}. In \cite{zarova05}, a detailed microscopic derivation is given, starting from the current transport equations in the resonator-wire system. The outcome is a nonlinearity similar to that postulated phenomenologically in \cite{zarov03}. We adopt essentially the same model, described by the electric permeability $\epsilon$ and the magnetic permittivity $\mu$:
\bea
\label{epseq}\epsilon(E,E^\dagger)&\equiv&\epsilon\left(\vert E\vert^2\right)=\left(\epsilon_{D0}+\alpha\vert E\vert^2\right)\left(1-\frac{\omega_0^2}{\omega(\omega+\imath\gamma)}\right)\\
\label{mueq}\mu(H,H^\dagger)&\equiv&\mu\left(\vert H\vert^2\right)=1+\frac{F\omega^2}{\omega_{0NL}^2\left(\vert H\vert^2\right)-\omega^2+\imath\Gamma\omega},
\eea
with $\alpha=1$ or $\alpha=-1$ for self-focusing or self-defocusing nonlinearity, respectively. Frequency is denoted by $\omega$ and $\epsilon_{D0}$ is the linear part of the permittivity. By $F$, $\gamma$ and $\Gamma$ we denote the filling factor of the material and the electric and magnetic damping coefficients. The equations (\ref{epseq}-\ref{mueq}) allow us to model also the real (lossless) dielectric response by putting $\gamma=0$. For the magnetic field, the permittivity will in general stay complex even for $\Gamma=0$, as the nonlinear frequency of the resonator rings $\omega_{0NL}$ can always have a nonzero imaginary part. This frequency is related to the magnetic field through the relation ($X\equiv\omega_{0NL}/\omega_0$):
\be
\label{om0eq1}\vert H\vert^2=\alpha A^2\frac{(1-X^2)\left[\left(X^2-\Omega^2\right)^2+\Omega^2\gamma^2\right]}{X^6},
\ee
where $\Omega\equiv\omega/\omega_0$, $\omega_0$ is the eigenfrequency of the rings, and $A$ is a parameter which can be derived microscopically \cite{zarova05,zarov03,sadriv05}; for our purposes, it can be treated just as a phenomenological parameter. This cubic equation yields three branches for $\omega_{0NL}^2$. All these branches are physical and correspond to different possible nonlinear oscillations \cite{sadriv05}. Now the equations of motion are just the Maxwell equations in a medium described by (\ref{epseq}-\ref{mueq}), in the approximation of slowly-changing beam envelopes. We assume an elongated (cylindrical or parallelopipedal) slab of metamaterial, so we can employ the paraxial beam approximation (e.g. \cite{book}). The beam is initially collimated along the longitudinal axis $z$ and focuses or defocuses slowly in the transverse $x-y$ plane due to the nonlinearity of the material. The electric and magnetic field $\hat{E}(t;x,y,z),\hat{H}(t;x,y,z)$ are directed along the $z$-axis. From now on, the speed of light is put to unity $c=1$. All the steps in deriving the nonlinear-Schr\"{o}dinger-like equation are well known so we merely state the final result here, which is quite close to the equations used in \cite{kivstamm} in $1+1$ dimension, or the equations found in \cite{wen,tsitsas,walasik}. Full derivation can be found in Appendix \ref{appa}. The equations of motion turn out to be:
\bea
\label{eome}-\frac{\imath}{b}\partial_zE&=&\nabla_\perp^2E+\left[\omega^2\epsilon\left(\vert E\vert^2\right)\mu\left(\vert H\vert^2\right)-k^2\right]E-
\frac{\nabla_\perp\mu\left(\vert H\vert^2\right)}{\mu\left(\vert H\vert^2\right)}\nabla_\perp E-\imath\frac{\partial_z\mu\left(\vert H\vert^2\right)}{2\mu\left(\vert H\vert^2\right)}E\\
\label{eomh}-\frac{\imath}{b}\partial_zH&=&\nabla_\perp^2H+\left[\omega^2\epsilon\left(\vert E\vert^2\right)\mu\left(\vert H\vert^2\right)-k^2\right]H.
\eea
Here, $\nabla_\perp\equiv(\partial_x,\partial_y)$ is the nabla operator in the transverse plane, $k$ is the wavevector along the $z$-direction, $b$ is the characteristic propagation length along the $z$-axis. Equations of motion (\ref{eome}-\ref{eomh}) together with the equations (\ref{epseq}-\ref{om0eq1}) for the permittivities contain the following five parameters: $\epsilon_{D0},F,\Gamma,\gamma,\omega_0$. Realistic values for all the parameters are discussed in \cite{sadriv05}. The natural length scale of the model is dominated by the $1/\omega_0$ scale. Dimensional analysis of the terms on the right-hand side of (\ref{eome}) determines the length scale $b$ in (\ref{eome}-\ref{eomh}) as $b\sim 1/\omega_0$. Both in analytical and in numerical calculations, we express the transverse coordinates ($x,y$) in millimeters but the longitudinal coordinate $z$ is often stated in units of $b$. This is because the length scales of all patterns in the transverse plane are similar whereas the propagation lengths along $z$ can vary by an order of magnitude as $\gamma$ and $\Omega$ are varied, so it is more natural to express them in terms of the characteristic distance $b$.

\subsection{A field-theoretical model}

For some theoretical considerations it is useful to formulate a Lagrangian (gradient) model which captures the essential features of the equations of motion (\ref{eome}-\ref{eomh}). As it often happens in studies of complex nonlinear pattern-forming systems, we cannot easily write the original equations in such a form. Instead, we construct a field theory which yields equations of motion somewhat different from the original ones but which still give the same phenomenology, and are able to explain the results of numerical calculations with the equations (\ref{eome}-\ref{eomh}).

Let us think what such a field theory would look like. The terms with the gradient of magnetic permittivity obviously introduce dissipation, which physically originates from the losses in the inductive rings of the metamaterial. In general, dissipative systems do not have a Lagrangian, although a number of generalized Lagrangian approaches exist for dissipative systems: either with more general functional forms of the Lagrangian, or with a dissipative function in addition to the Lagrangian, or with extra degrees of freedom \cite{disslag1,disslag2}. We will take the first, most conventional of the three approaches: we will consider a conventional Lagrangian (no dissipative function, no extra degrees of freedom) which gives slightly generalized equations of motion compared to (\ref{eome}-\ref{eomh}), with dissipative terms for both electric and magnetic fields coming from the complex terms in the effective potential. The effective action reads:
\bea
\nonumber\mathcal{L}&=&\mathcal{L}_E+\mathcal{L}_H\\
\nonumber\mathcal{L}_E&=&\frac{\imath}{2\mu\left(\vert H\vert^2\right)}\left(E\partial_zE^\dagger-E^\dagger\partial_zE\right)+\frac{\vert\nabla_\perp E\vert^2}{\mu\left(\vert H\vert^2\right)}+\frac{k^2\vert E\vert^2}{\mu\left(\vert H\vert^2\right)}-\omega^2\epsilon\left(\vert E\vert^2\right)\vert E\vert^2\\
\label{lag}\mathcal{L}_H&=&\frac{\imath}{2\epsilon\left(\vert E\vert^2\right)}\left(H\partial_zH^\dagger-H^\dagger\partial_zH\right)+\frac{\vert\nabla_\perp H\vert^2}{\epsilon\left(\vert E\vert^2\right)}+\frac{k^2\vert H\vert^2}{\epsilon\left(\vert E\vert^2\right)}-\omega^2\int_0^{HH^\dagger}dx\mu(x).
\eea
The last term in $\mathcal{L}_E$ equals $-\omega^2\int_0^{EE^\dagger}dx\epsilon(x)$, analogously to the corresponding term in $\mathcal{L}_H$, but since $\epsilon$ is polynomial in $E^\dagger E$, the integral can be solved explicitly. Now (\ref{lag}) gives the equations of motion:
\bea
\label{eomlae}-\frac{\imath}{b}\partial_zE&=&\nabla_\perp^2E+\left[\epsilon\left(\vert E\vert^2\right)\mu\left(\vert H\vert^2\right)-k^2\right]E-\frac{\imath\partial_z\mu\left(\vert H\vert^2\right)}{\mu\left(\vert H\vert^2\right)}E-\frac{\nabla_\perp\mu\left(\vert H\vert^2\right)}{\mu\left(\vert H\vert^2\right)}\nabla_\perp E-\Phi_H\\
\label{eomlah}-\frac{\imath}{b}\partial_zH&=&\nabla_\perp^2H+\left[\epsilon\left(\vert E\vert^2\right)\mu\left(\vert H\vert^2\right)-k^2\right]H- \frac{\imath\partial_z\epsilon\left(\vert E\vert^2\right)}{\epsilon\left(\vert E\vert^2\right)}H-\frac{\nabla_\perp\epsilon\left(\vert E\vert^2\right)}{\epsilon\left(\vert E\vert^2\right)}\nabla_\perp H-\Phi_E,
\eea
where $\Phi_{E,H}$ are related to the fluxes of the electric and magnetic field (prime denotes the derivative of $\epsilon$ and $\mu$ with respect to their arguments $E^\dagger E$ and $H^\dagger H$):
\bea
\label{fluxh}\Phi_H&=&
\frac{\epsilon'\left(\vert E\vert^2\right)}{\epsilon^2\left(\vert E\vert^2\right)}\left(\frac{\imath}{2}\left(H\partial_zH^\dagger-H^\dagger\partial_zH\right)+\vert\nabla_\perp H\vert^2+k^2\vert H\vert^2\right)\\
\label{fluxe}\Phi_E&=&
\frac{\mu'\left(\vert H\vert^2\right)}{\mu^2\left(\vert H\vert^2\right)}\left(\frac{\imath}{2}\left(E\partial_zE^\dagger-E^\dagger\partial_zE\right)+\vert\nabla_\perp E\vert^2+k^2\vert E\vert^2\right).
\eea
These are the extra terms compared to the physical equations (\ref{eome}-\ref{eomh}).\footnote{The dissipative term proportional to $\nabla_\perp H$ in (\ref{eomlah}) is also absent in the original equations, but that one is easy to interpret: we make both $\mathcal{L}_E$ and $\mathcal{L}_H$ complex, so both fields have dissipative dynamics.} Inserting $\partial_zE^\pm$ from the equations of motion (\ref{eomlae}-\ref{eomlah}) into the above we derive:
\be
\label{eomflux}\Phi_E=\frac{\mu'}{\mu}\nabla_\perp\left(\frac{E\nabla_\perp E^\dagger-E^\dagger\nabla_\perp E}{\mu}\right).
\ee
and analogously for $\Phi_H$, with $\epsilon\leftrightarrow\mu,E\leftrightarrow H$. This term is proportional to a total derivative, and is therefore related to the flux $(E\nabla_\perp E^\dagger-E^\dagger\nabla_\perp E)/\mu$. For slowly-changing $\epsilon$ and $\mu$, which is often the case in our system (i.e., for $\epsilon^\prime,\mu^\prime\ll\epsilon,\mu$), this term is small, which partly justifies the choice (\ref{lag}) for the Lagrangian. But the ultimate justification, as it frequently happens, is that \emph{a posteriori} we will find that this model is able to explain the features observed in the numerics. Therefore we will not try to interpret the term (\ref{eomflux}) in detail.

\section{Geometry and stability of vortices}

We will now sum up our numerical results which demonstrate the breaking of the circular symmetry of the vortex beams and their decay during the propagation. We always start from a Gaussian input beam with a topological charge $Q$, of the form $E(r,\phi;z=0)=E_0\times e^{-r^2/2\sigma^2} e^{iQ \phi}$ and analogously for the magnetic field, with amplitude $H_0$ but with the same vortex charge $Q$. Therefore, we always give an exact vortex as an input. The parameters of the model were chosen so that the permittivities $\epsilon$ and $\mu$, given by \eqref{epseq} and \eqref{mueq} respectively, are of order unity. This serves to limit the dissipative effects, so that the propagation along the longitudinal direction can be clearly observed. Same phenomena are found for arbitrary values of $\epsilon$ and $\mu$ but on a different length scale. We do not aim at a stamp-collecting exhaustive description of patterns for all possible parameter values, so we will focus on just a few relevant cases. We are mainly interested in left-handed materials ($\epsilon,\mu<0$) and how they compare to right-handed ones, so for the dielectric constant we always choose the self-defocusing Kerr nonlinearity ($\alpha=-1$) with a linear part $\epsilon_{D0} = 12.8$, which has both a left-handed and a right-handed regime. To check the effects of dissipation, we either adopt $\gamma=0$ in (\ref{epseq}), i.e., the lossless case, or we tune $\gamma$ so that $\omega_0^2/(\omega^2+\imath\gamma\omega)=1/2$. In other words, we impose either $\Im\epsilon=0$ or $\Im\epsilon=\Re\epsilon$. This is for simplicity and to avoid probing a huge parameter space for all possible $\gamma$ values; from now on we will call these cases simply lossless $\epsilon$ and dissipative $\epsilon$. The filling factor is $F=0.4$ and the magnetic dampening coefficient is $\Gamma=10^9$ Hz; these values are kept fixed in all calculations. Numerical calculations are performed with an operator split algorithm described in detail in the Appendices of \cite{nasstari}.

\begin{figure}[H]
\centering
(a)\includegraphics[width=.3\textwidth]{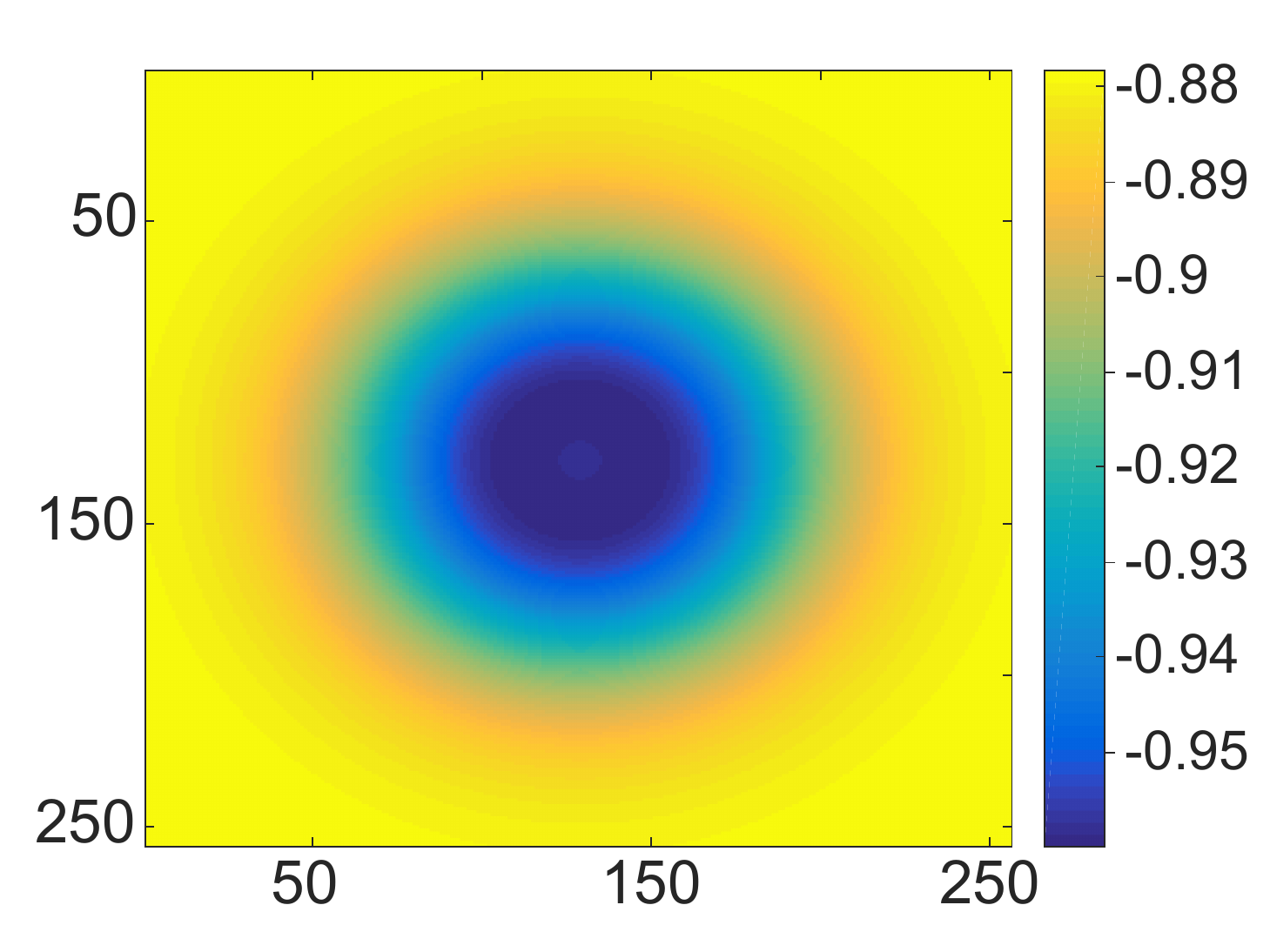}
(b)\includegraphics[width=.3\textwidth]{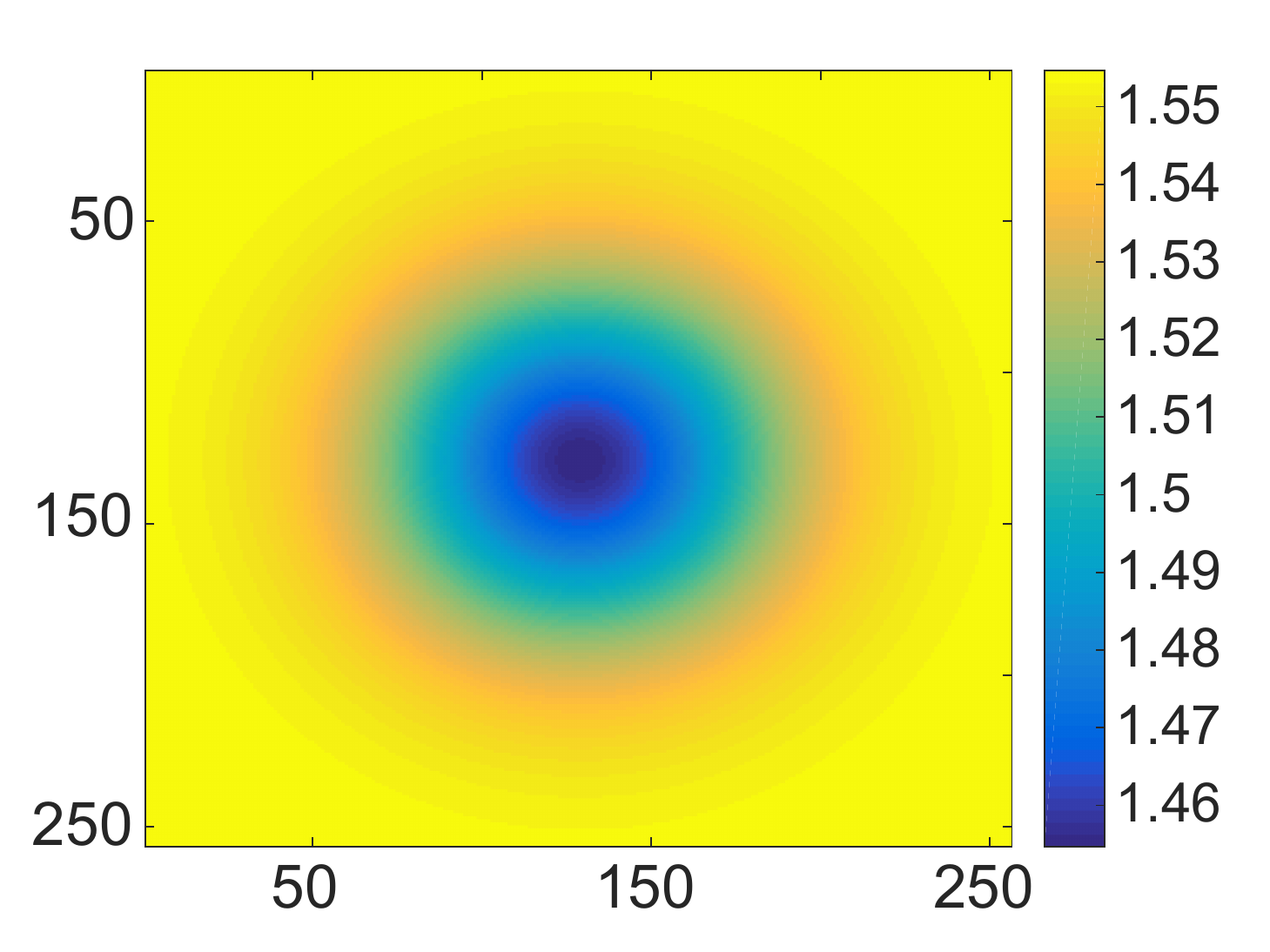}
(c)\includegraphics[width=.3\textwidth]{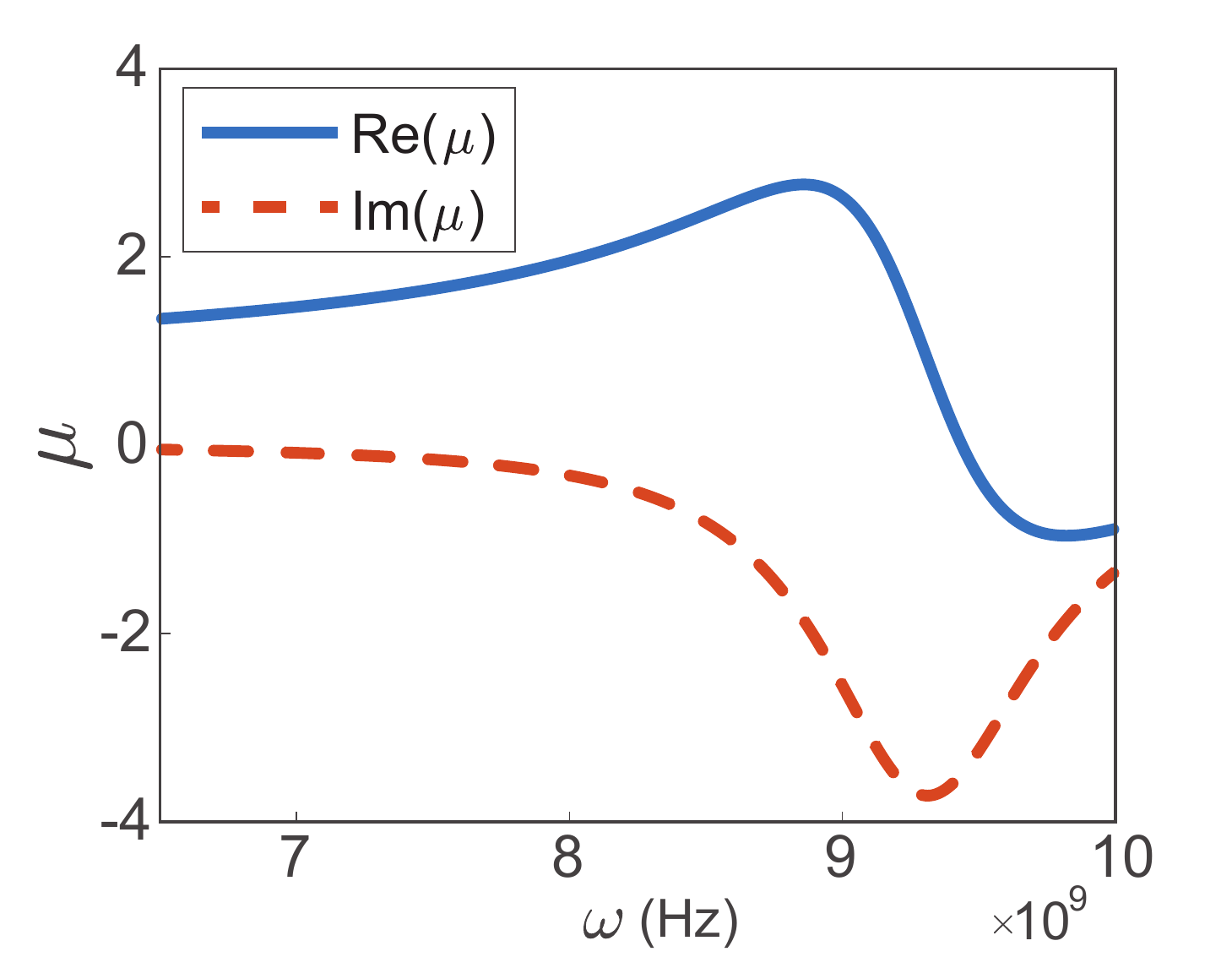}
\caption{The radial profile of $\mu$ for a left-handed medium (a) and a right-handed medium (b), for a vortex of charge $Q=1$. The profiles are radially symmetric in accordance with the fact that $\mu$ depends strictly on the magnitude of the magnetic field vector $\vert H \vert^2$. The real (blue) and imaginary (red) values of the complex permeability $\mu$ vs the frequency of the beam $\omega$ are displayed in (c). For frequencies higher than the eigenfrequency of the resonator rings $\omega_0$, the real part of the permeability is negative, essentially yielding a left-handed medium. The figure is made for dissipative $\epsilon$; for lossless $\epsilon$ the behavior is similar.}
\label{material}
\end{figure}

The nonlinear frequency of the oscillator rings is obtained as a solution to \eqref{om0eq1}. Of the three branches of the solution, we take the one that yields a negative real value of $\mu$ for $\omega>\omega_0$ (Fig.~\ref{material}). We have freely taken $\omega = 9.8 \times 10^9$ Hz to represent a left-handed medium, and $\omega = 7.0 \times 10^9$ Hz to represent a right-handed medium. The transverse profiles are displayed in Fig.~\ref{material}. We see there is a well-defined left-handed regime.

\begin{figure}[H]
 \centering
(a)
\includegraphics[width=.2\textwidth]{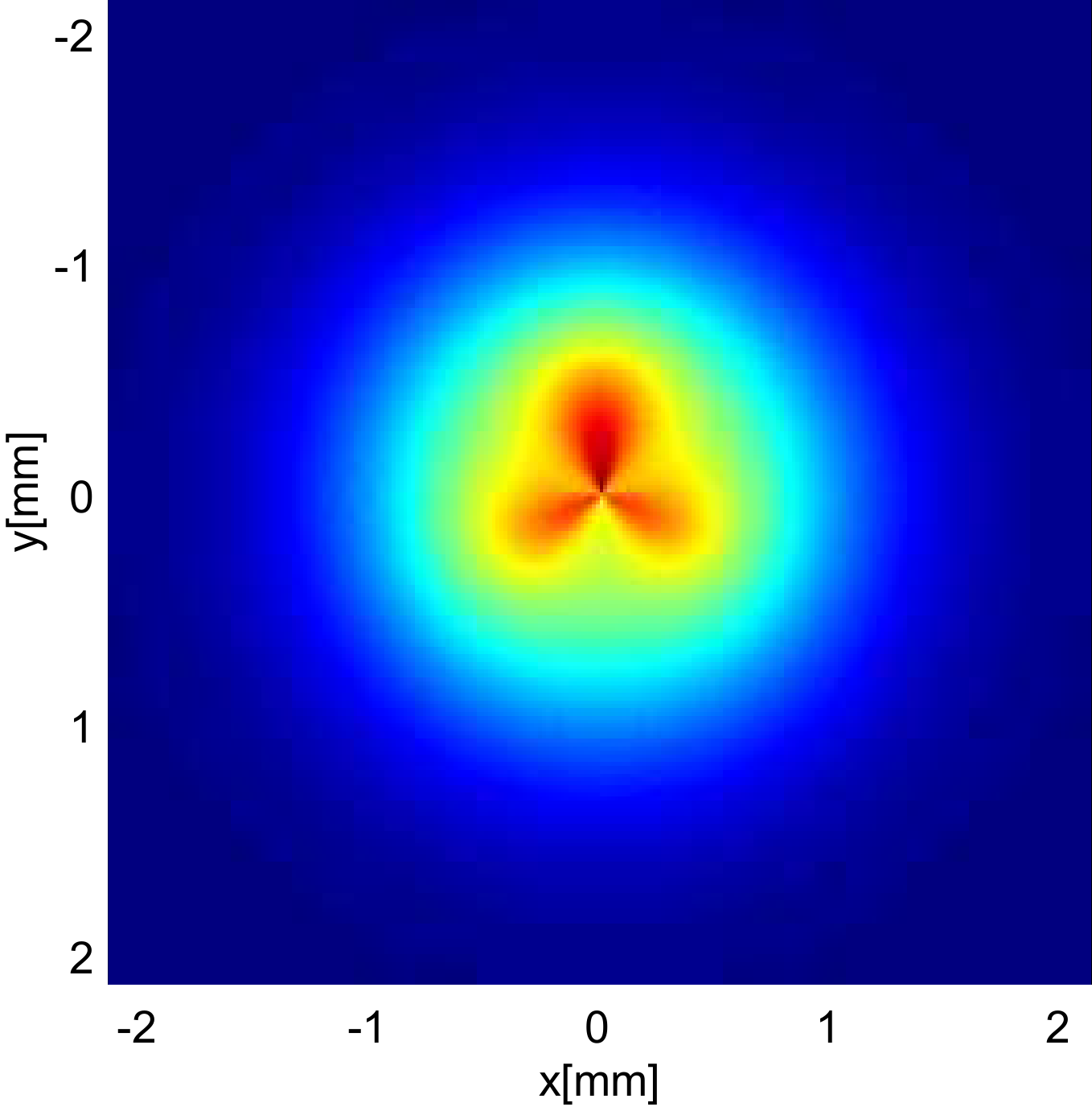}
\includegraphics[width=.2\textwidth]{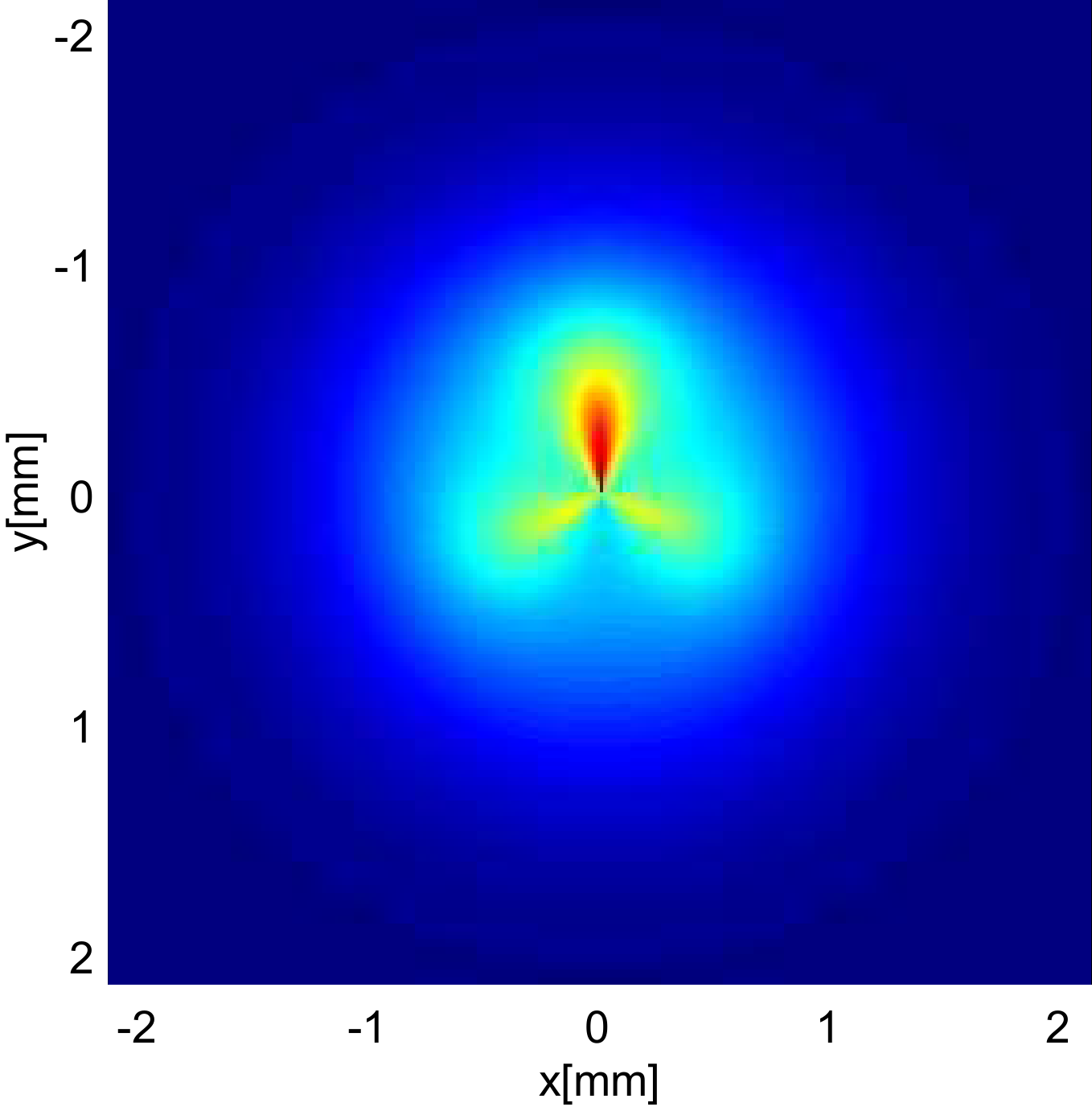}
\includegraphics[width=.2\textwidth]{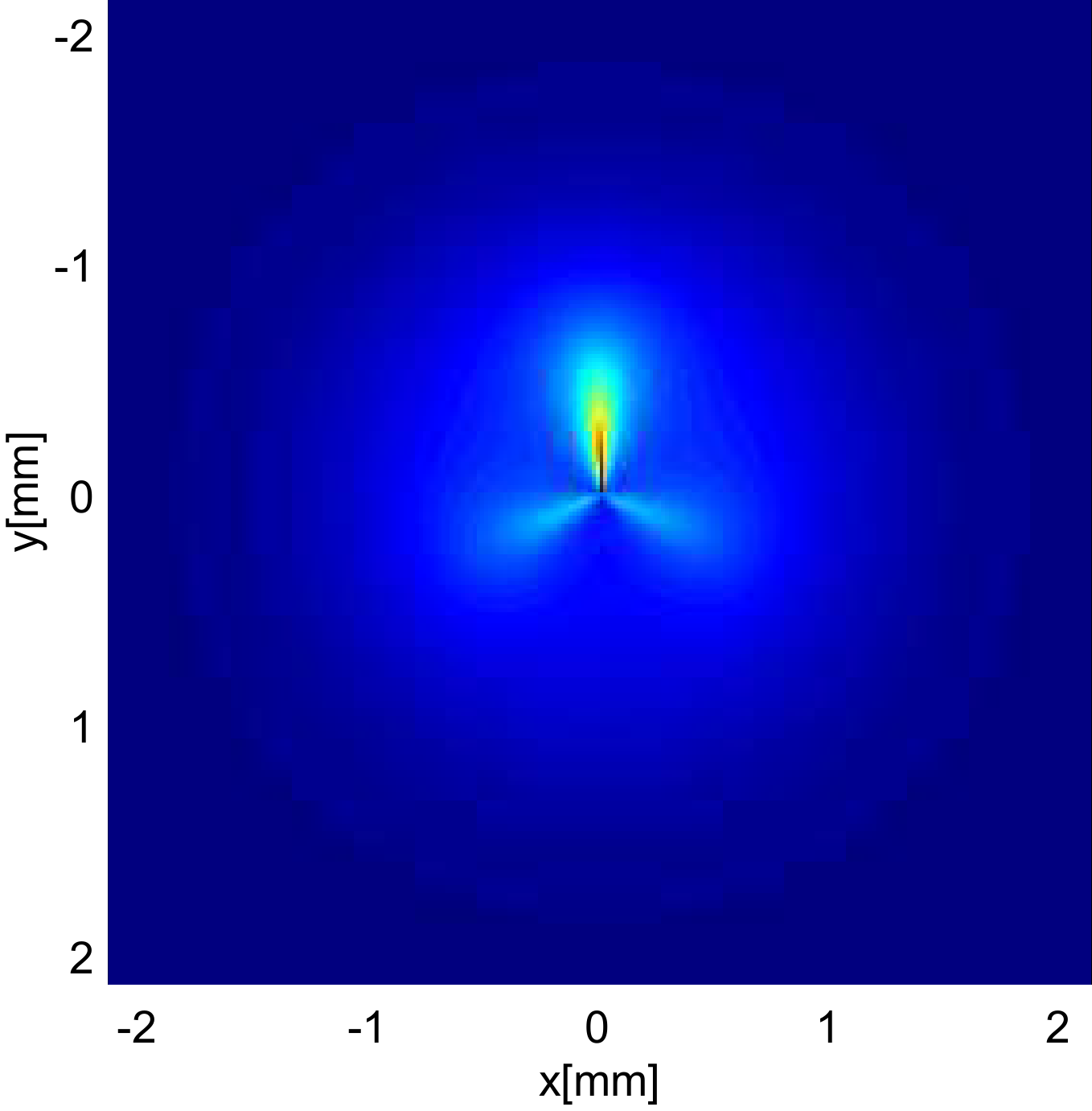}
\includegraphics[width=.2\textwidth]{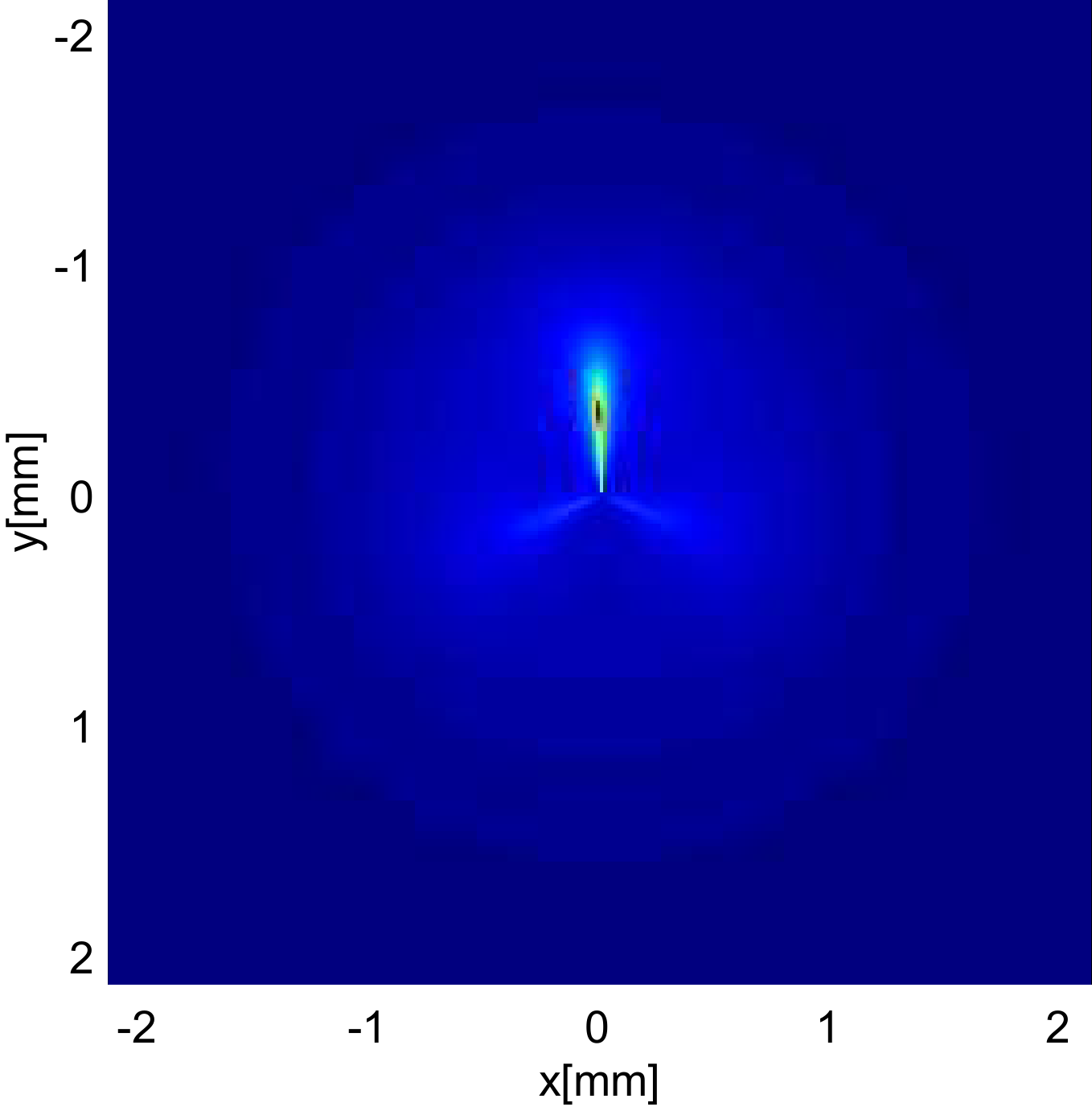}

(b)
\includegraphics[width=.2\textwidth]{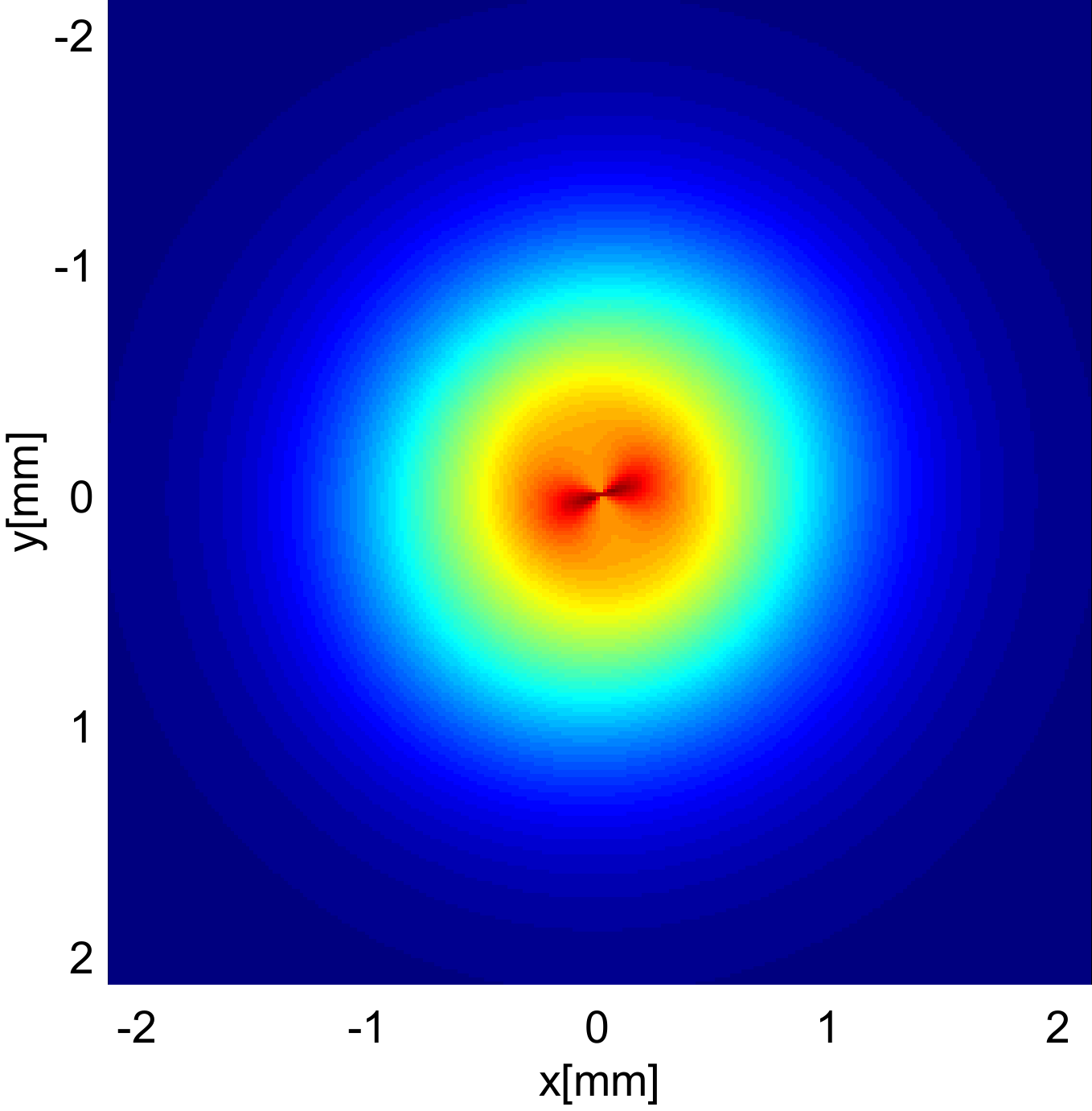}
\includegraphics[width=.2\textwidth]{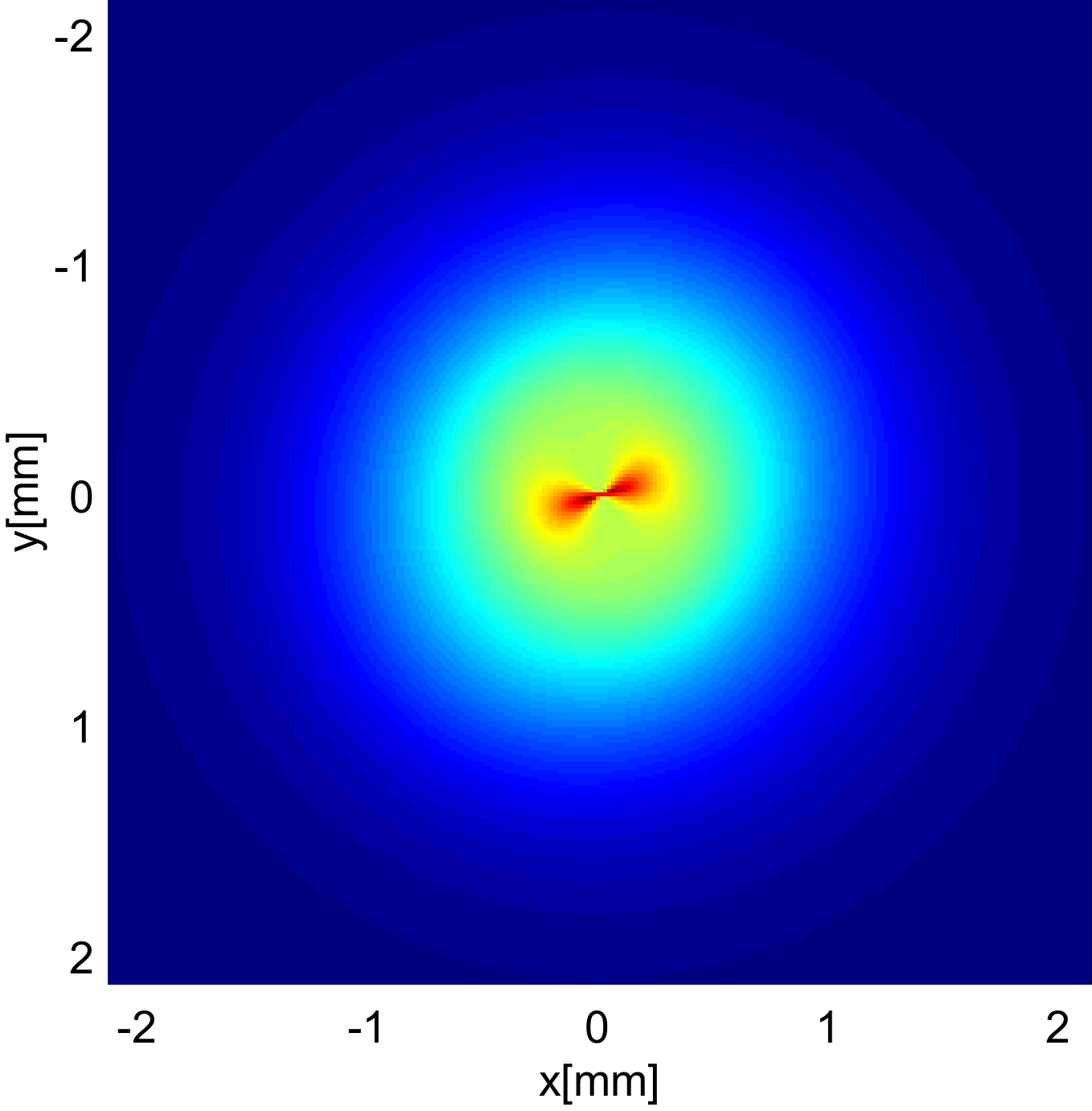}
\includegraphics[width=.2\textwidth]{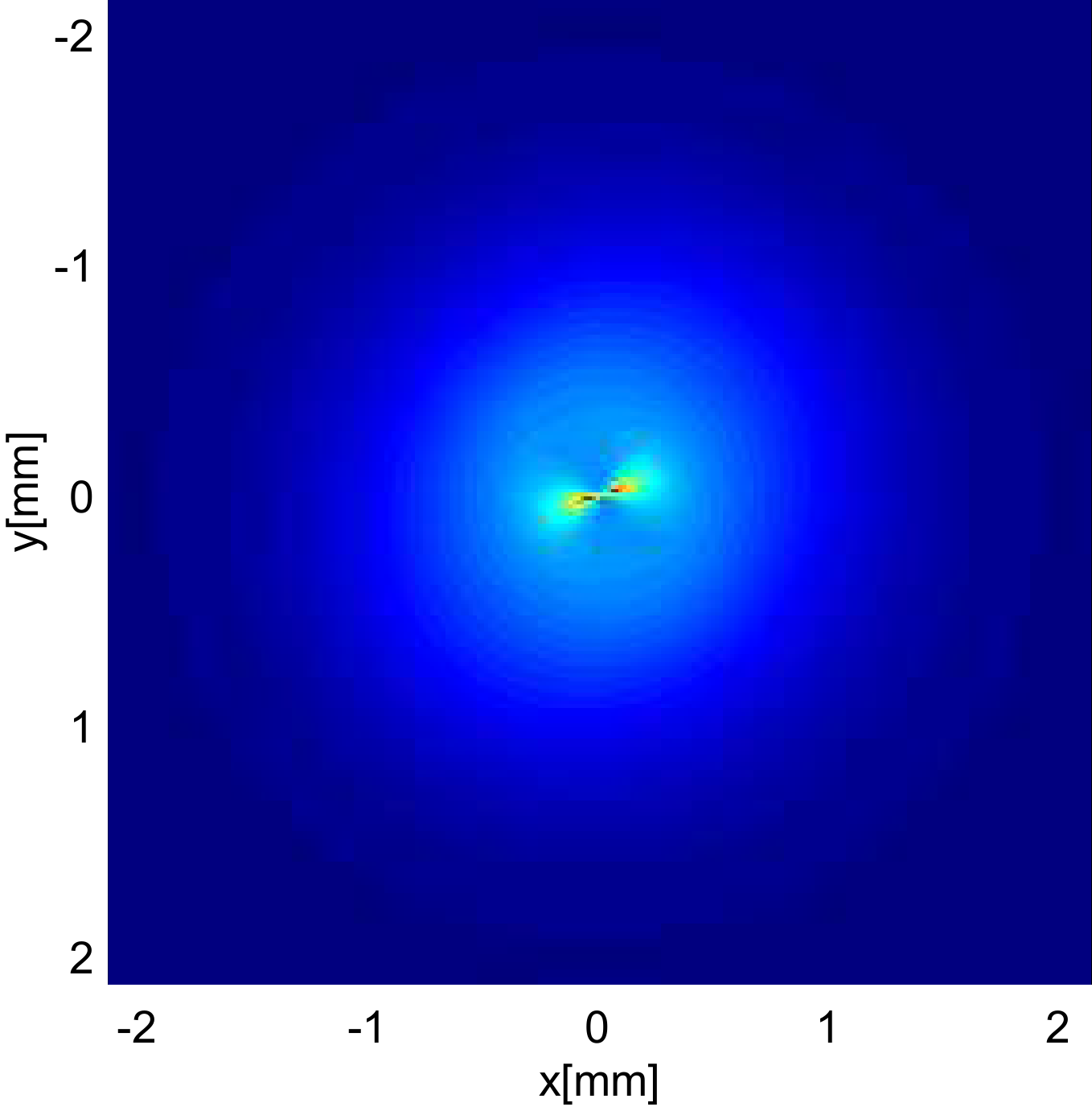}
\includegraphics[width=.2\textwidth]{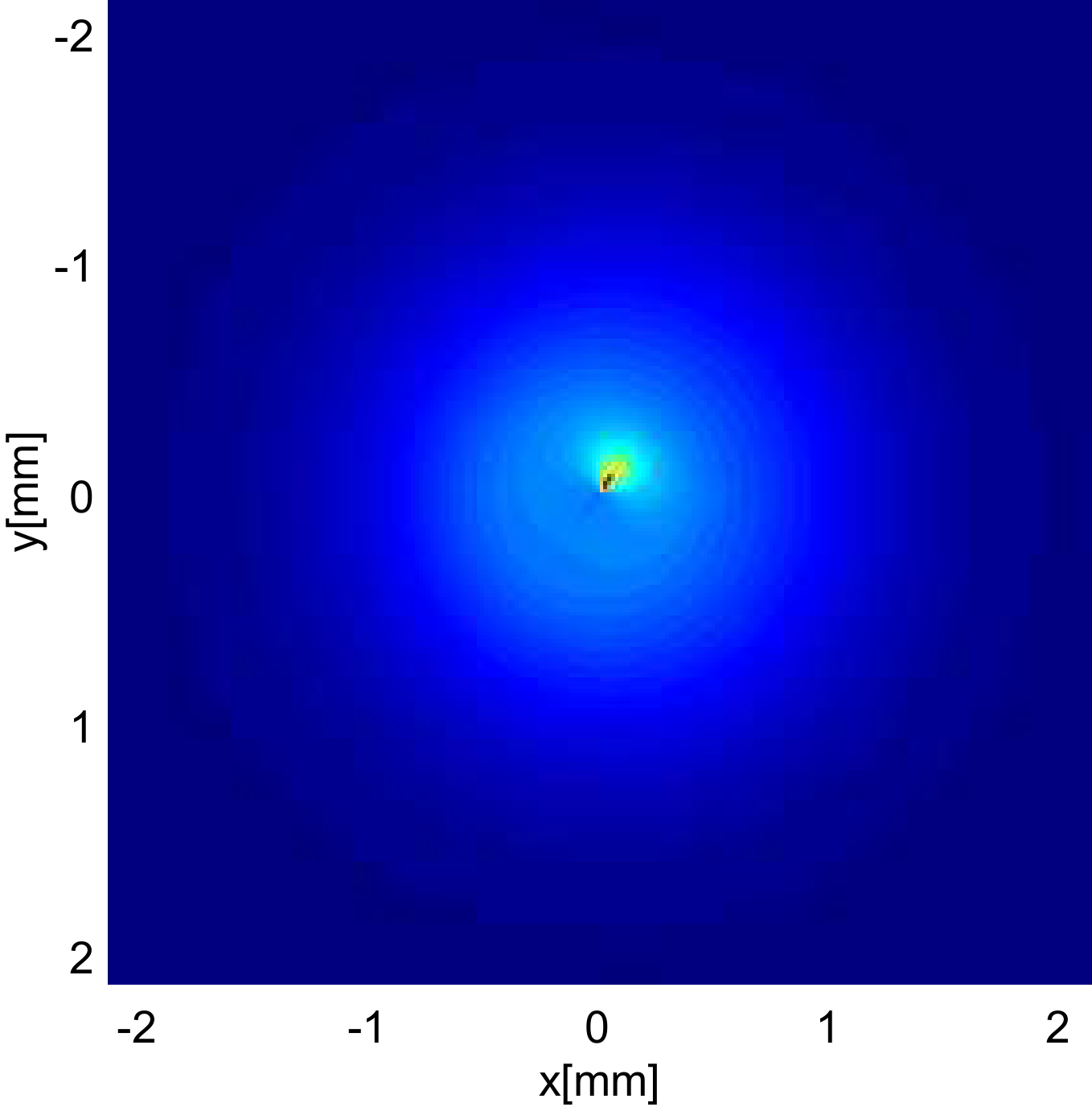}

(c)
\includegraphics[width=.2\textwidth]{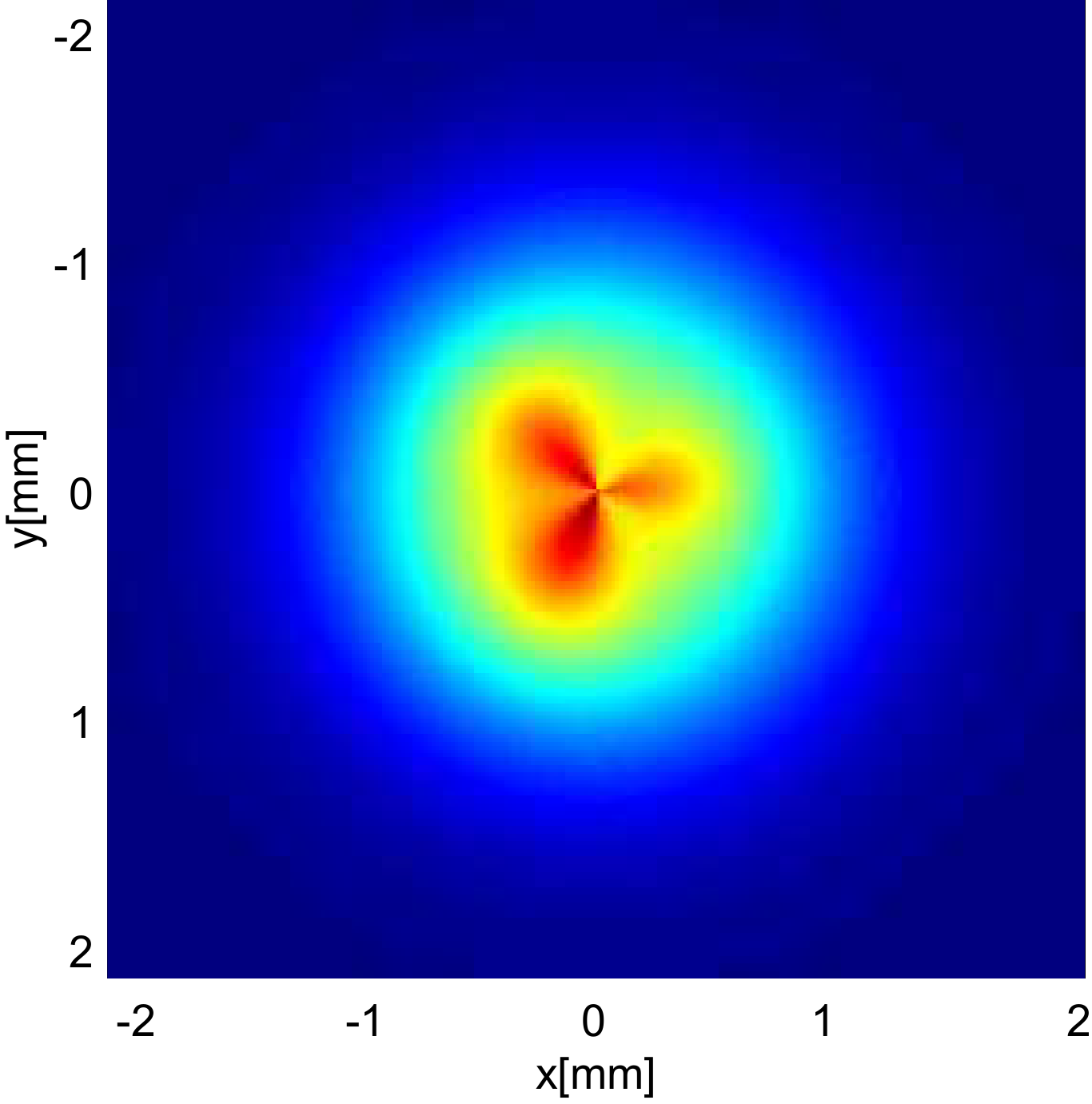}
\includegraphics[width=.2\textwidth]{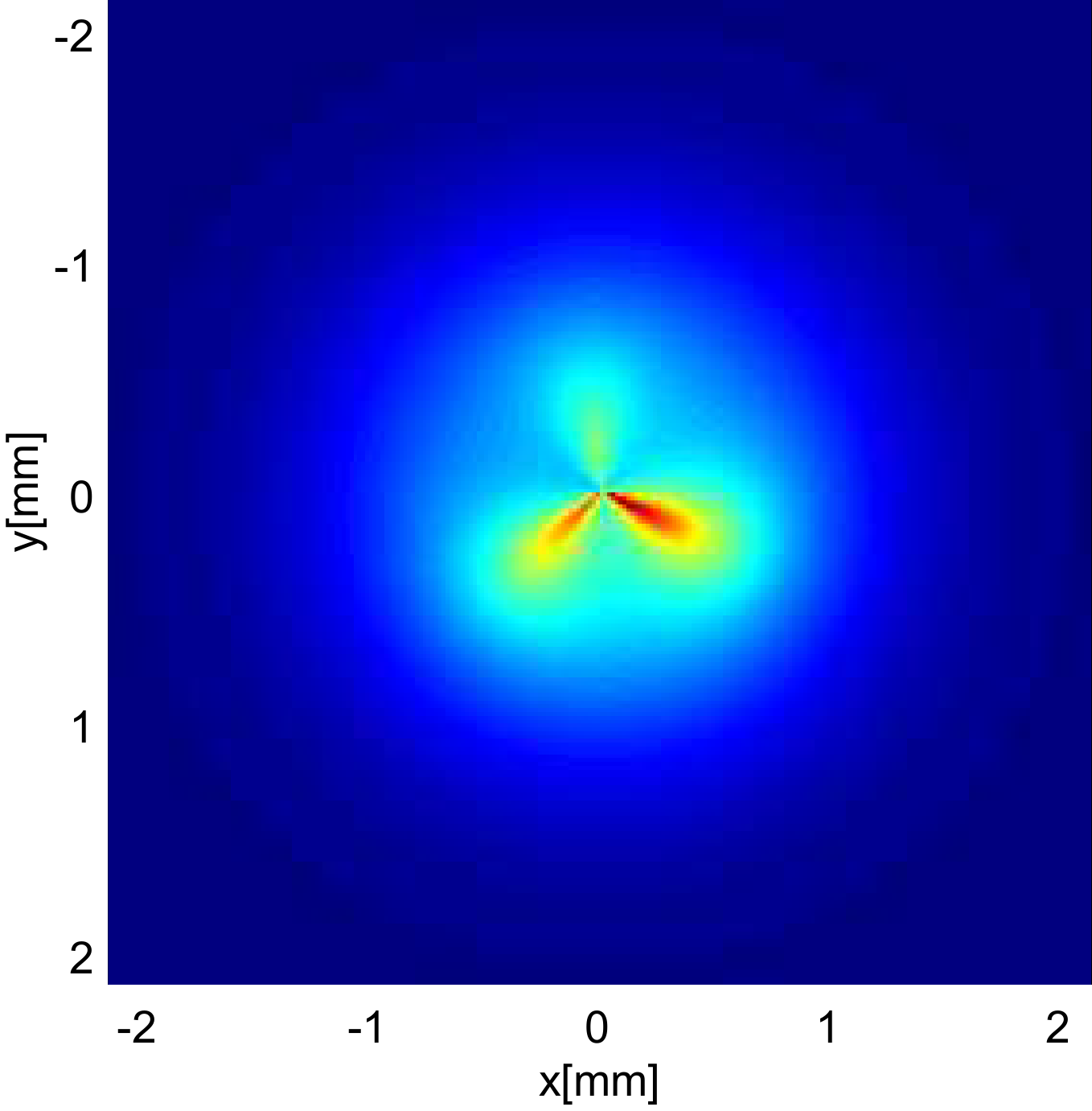}
\includegraphics[width=.2\textwidth]{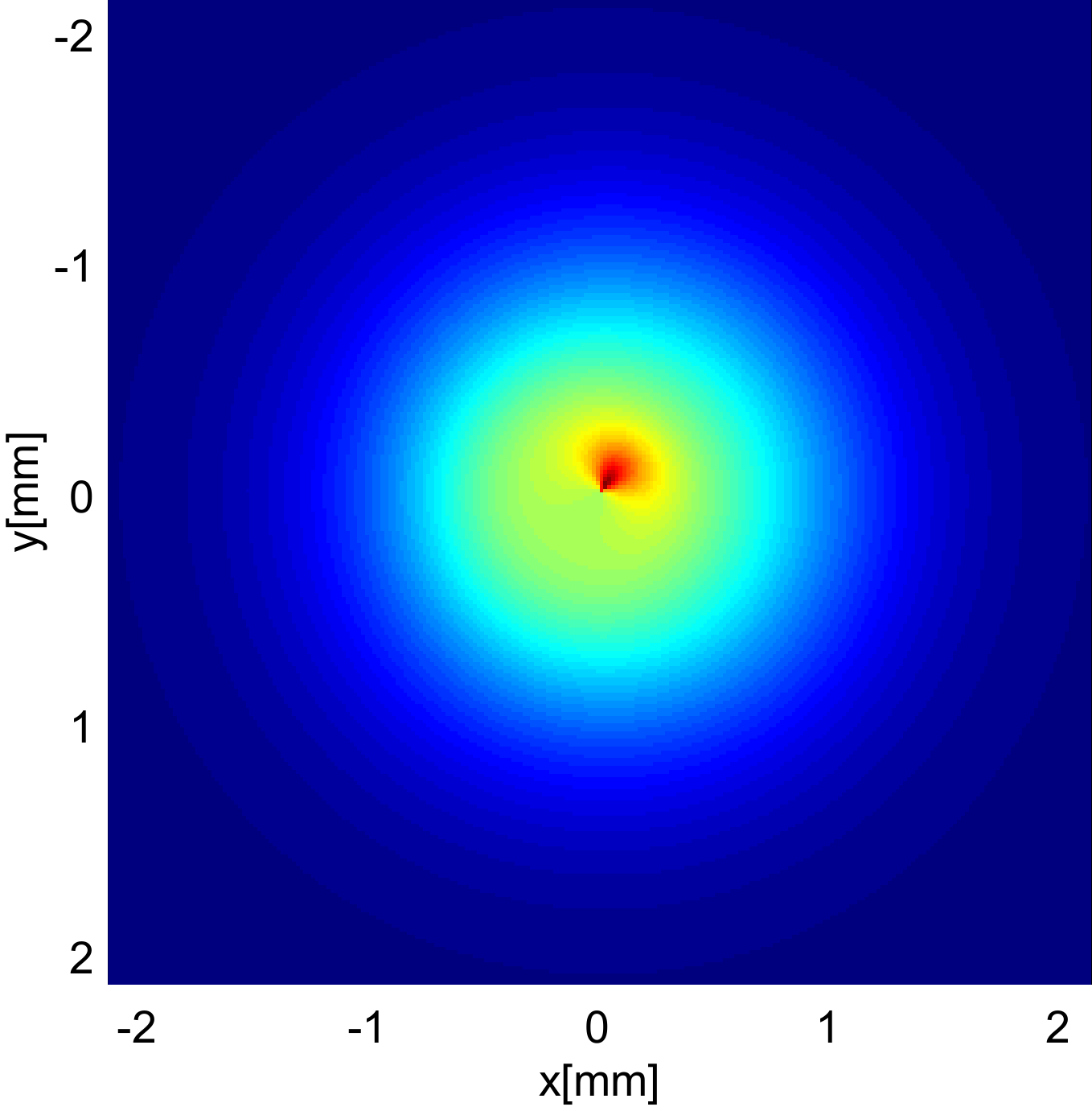}
\includegraphics[width=.2\textwidth]{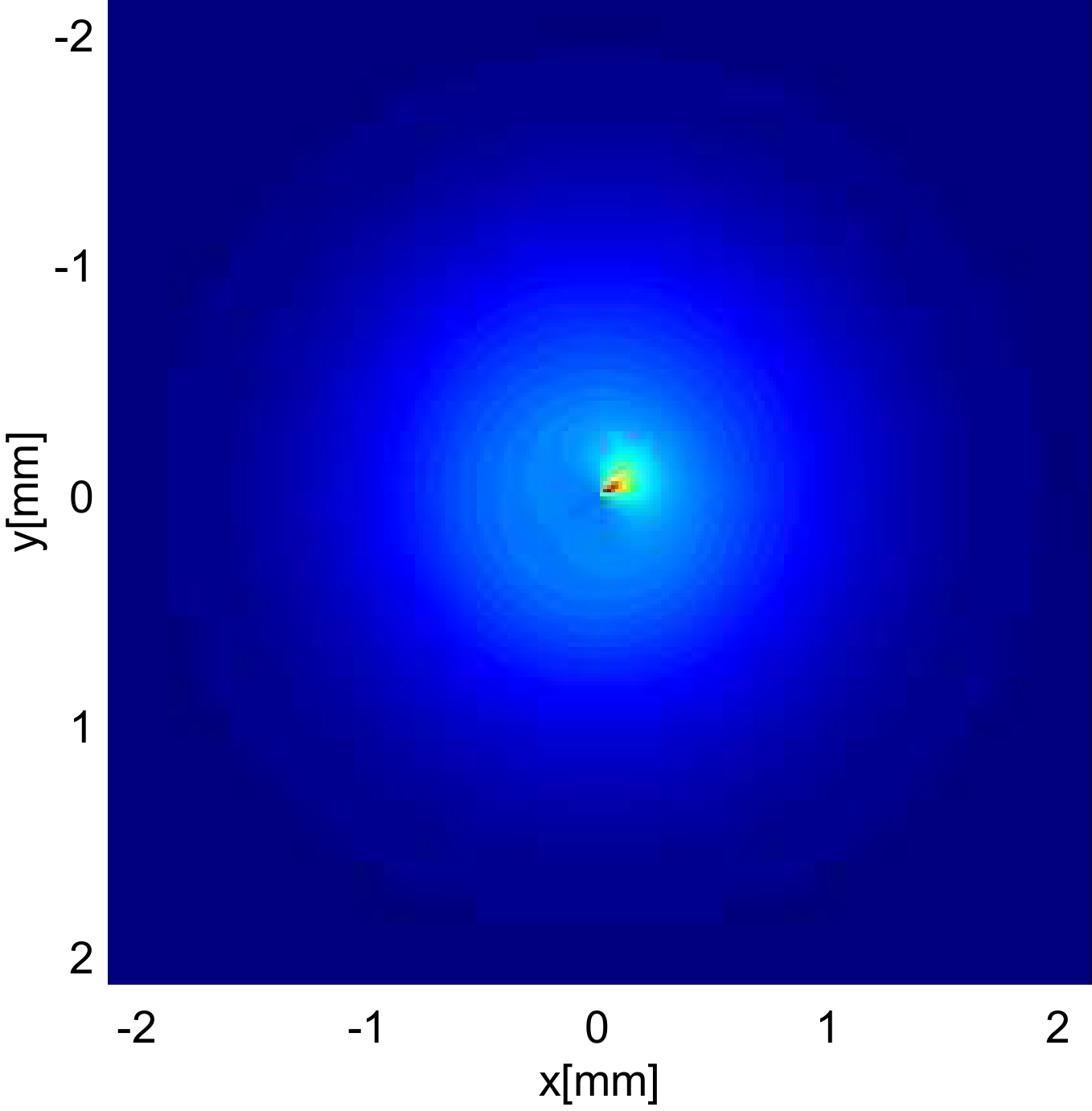}

(d)
\includegraphics[width=.2\textwidth]{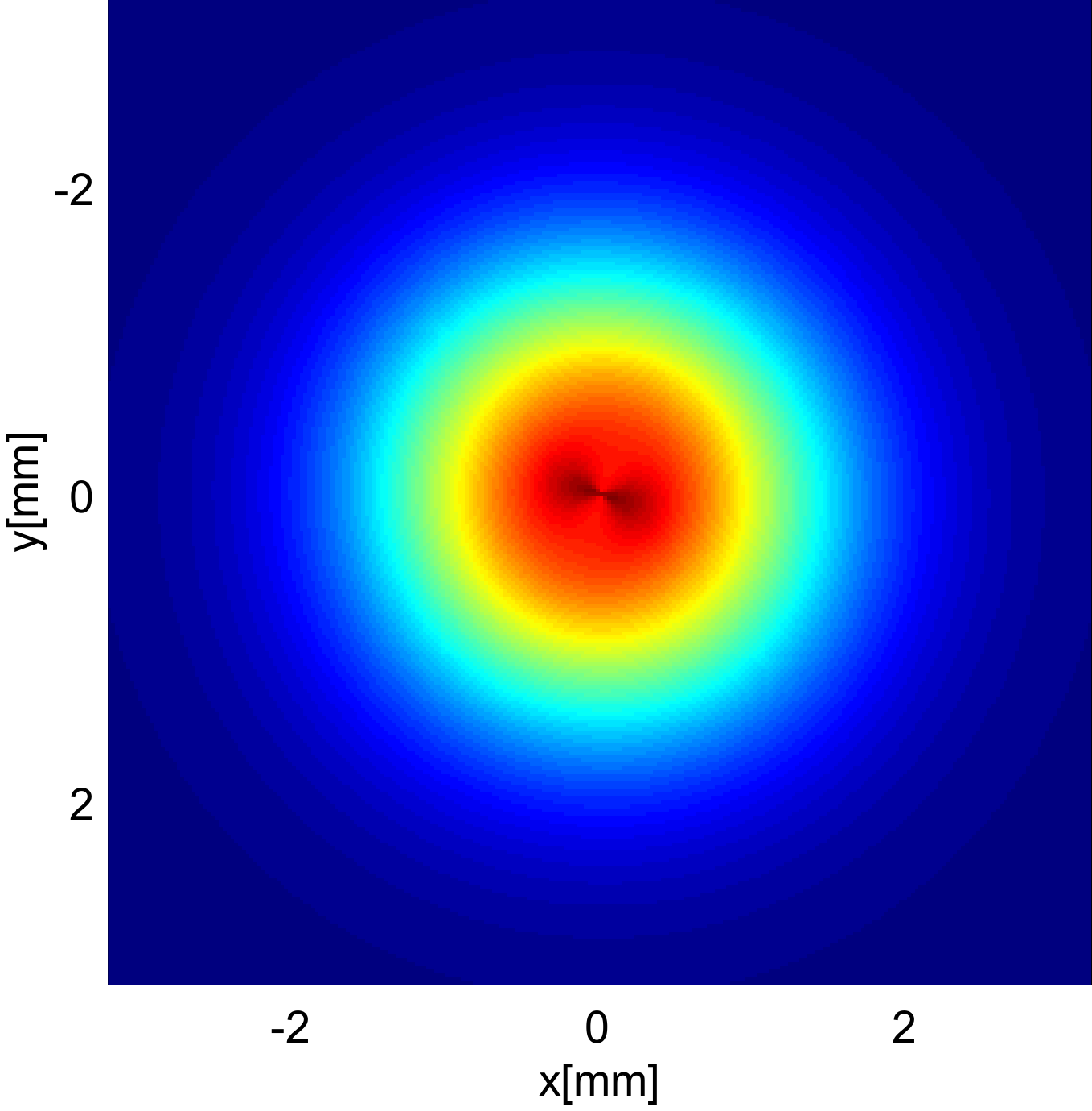}
\includegraphics[width=.2\textwidth]{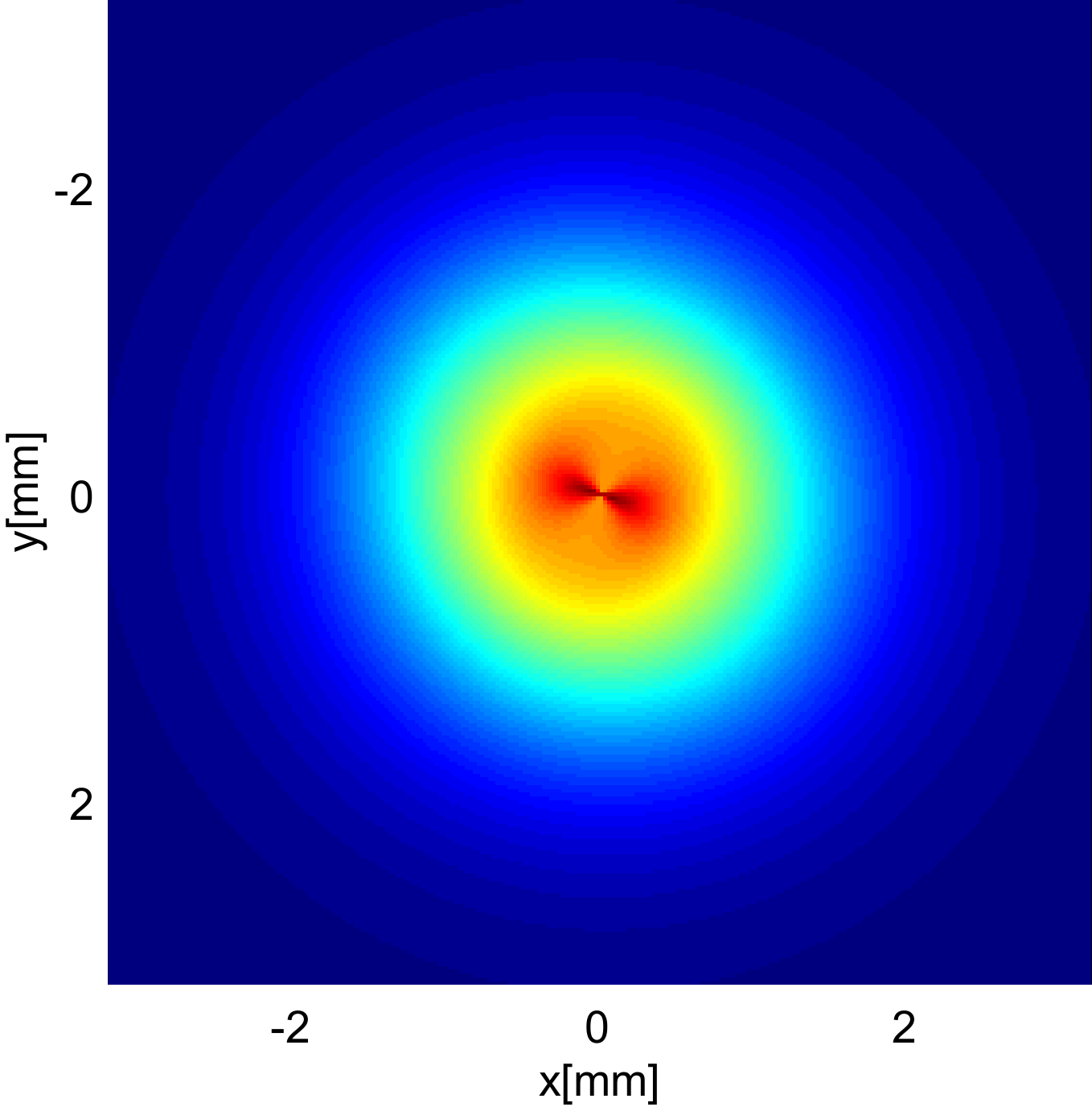}
\includegraphics[width=.2\textwidth]{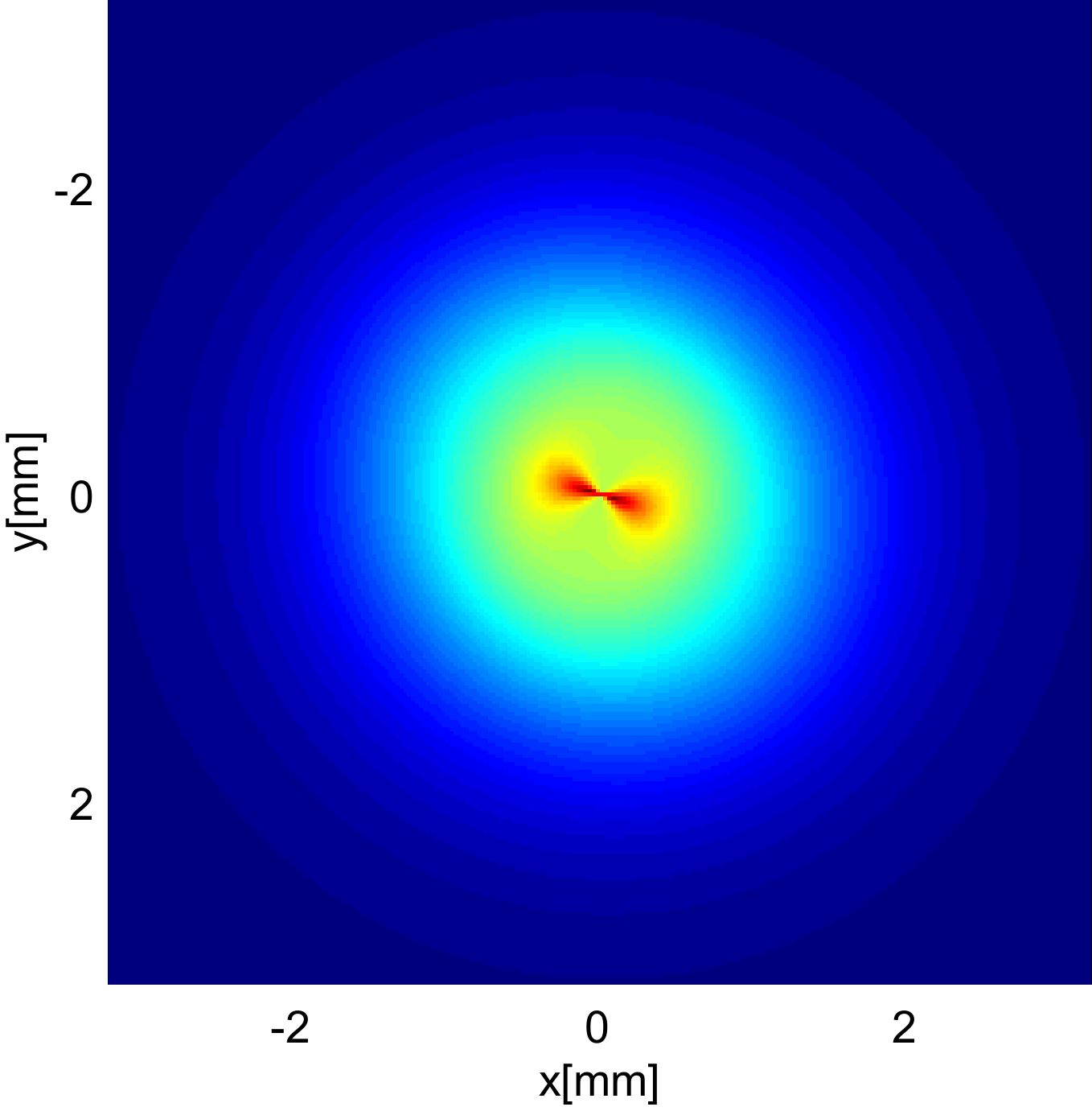}
\includegraphics[width=.2\textwidth]{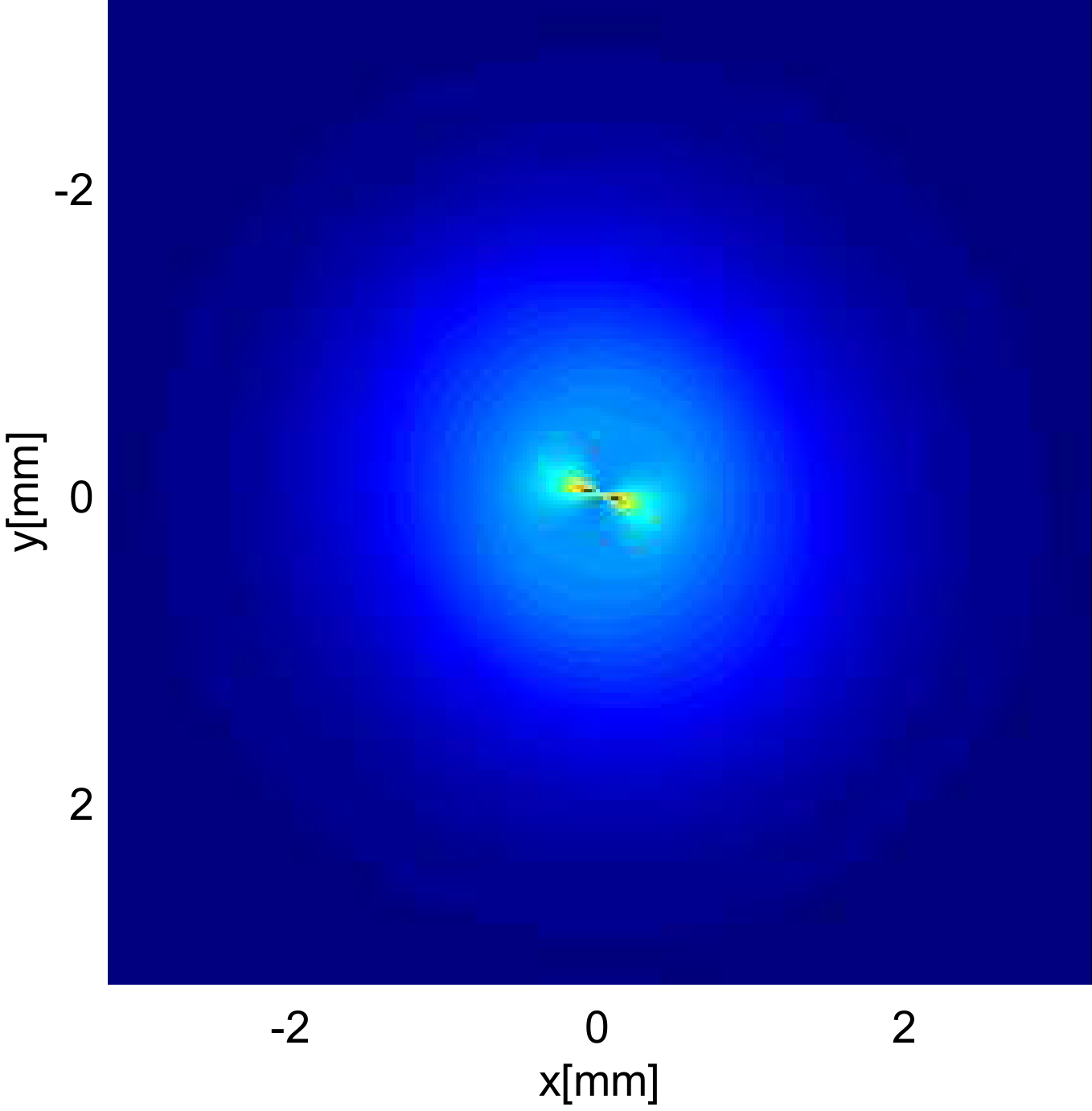}

\caption{The patterns for a $Q=1$ vortex, in left/right handed dissipative medium (a/b), and in left/right-handed lossless medium (c/d), at longitudinal slices $z=2b,4b,6b,8b$, showing the $C_{3Q}$, $C_{2Q}$, $C_{3Q}/C_Q$, $C_{2Q}$. The remaining parameters are defined in the text at the beginning of this section.}
\label{pattvoc}
\end{figure}

\begin{figure}[H]
\centering
\includegraphics[width=.5\textwidth]{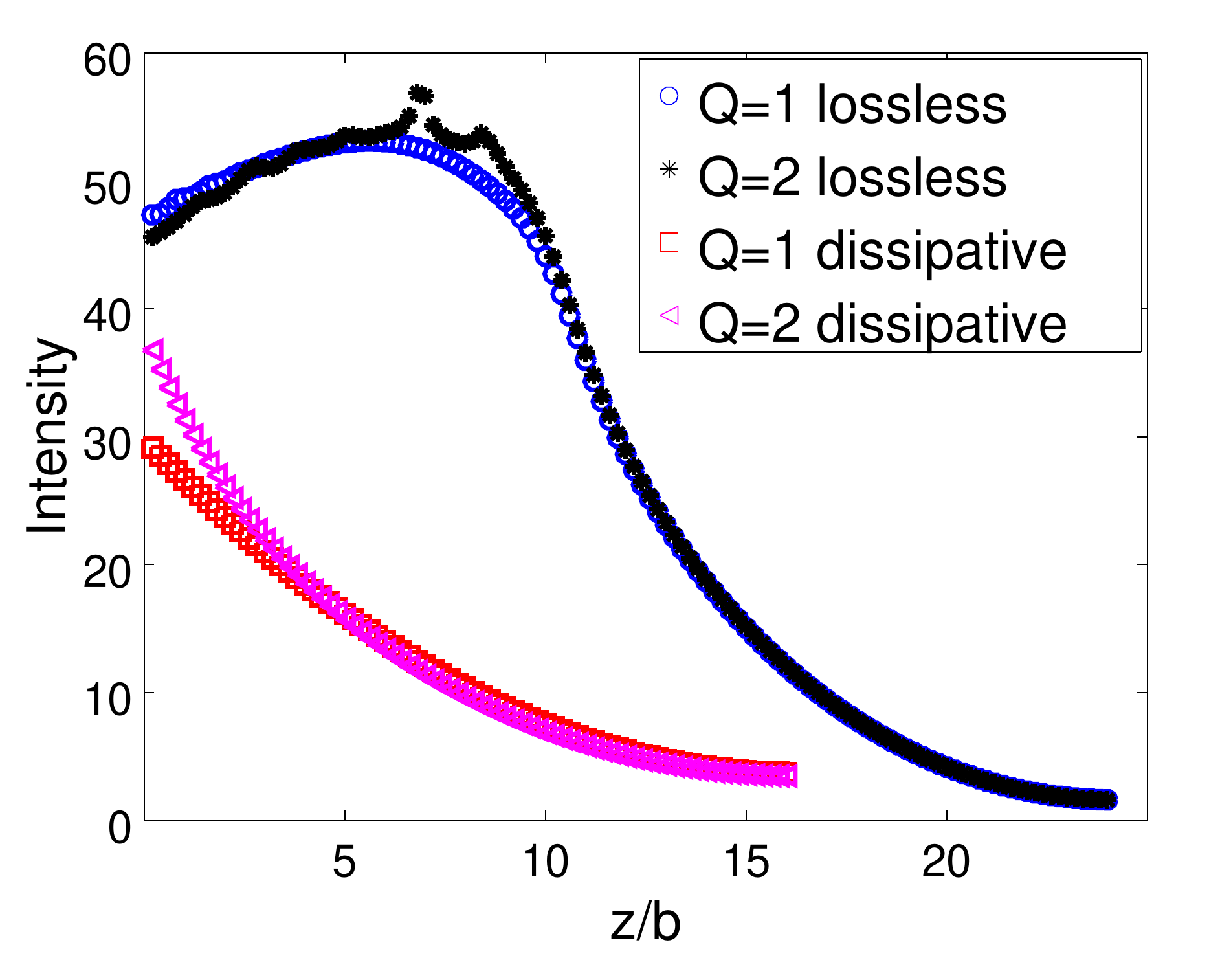}
\caption{Decay of the total intensity $I=\int dx\int dy (E^2+H^2)$ in computational units for $Q=1,2$, for a lossless (blue circles, black stars) and dissipative (red squares, magenta triangles) left-handed material. At early times the behavior is complicated and non-universal but at late times it collapses to an exponential curve. This is expected when loss through radiation dominates. The oscillatory features of the $Q=2$ lossless case (black) are likely due to finite numerical resolution.}
\label{int-decay}
\end{figure}

Now we discuss the transverse intensity profile for different initial beam configurations, with vortex input beams as explained in the beginning. We observe the following features:
\begin{enumerate}
 \item[1.] Circular symmetry of the vortex input always breaks down to a discrete group.
 \begin{enumerate}
 \item[(a)] In a dissipative left-handed medium, the discrete symmetry group for a vortex of charge $Q$ is $C_{3Q}$, before breaking down to simple $C_2$ axial symmetry at longer distances -- Fig~\ref{pattvoc}(a).
 \item[(b)] In a dissipative right-handed medium, the discrete symmetry group for a vortex of charge $Q$ is $C_{2Q}$, before breaking to $C_Q$ and then to $C_2$ axial symmetry at longer distances -- Fig.~\ref{pattvoc}(b).
 \item[(c)] In a lossless left-handed medium, the discrete symmetry group for a vortex of charge $Q$ is $C_{3Q}$ for very short distances, before quickly breaking down to $C_Q$ and finally $C_2$ -- Fig.~\ref{pattvoc}(c).
 \item[(d)] In a lossless right-handed medium, the discrete symmetry group for a vortex of charge $Q$ is $C_{2Q}$, before breaking to simple $C_2$ axial symmetry at longer distances -- Fig.~\ref{pattvoc}(d).
 \end{enumerate}
 \item[2.] Vortices decay approximately exponentially as they propagate along the longitudinal axis. Fig.~\ref{int-decay} shows the intensity of the beam across the $z$-axis, for various regimes. At early $z$ values, total intensity may behave non-monotonically and non-universally but on longer scales it decays exponentially. For different charges, the intensity curves collapse to a unique exponential function at large $z$. As could be expected, lossless and dissipative cases differ somewhat and collapse to different curves.
\end{enumerate}
The bottom line is that there is a vocabulary of patterns with $C_Q,C_{2Q},C_{3Q}$ symmetries. One of them  dominates in each case (left/right handed, dissipative/lossless) but at longer propagation distances the symmetry can change,
before the intensity drops to near-zero from dissipation. The final stadium of $C_2$ symmetry is only seen at very low intensities, so it might be practically unobservable in experiment; that is why we say the vocabulary only has
three possible patterns, excluding $C_2$.

\begin{figure}[H]
 \centering
(a)
\includegraphics[width=.2\textwidth]{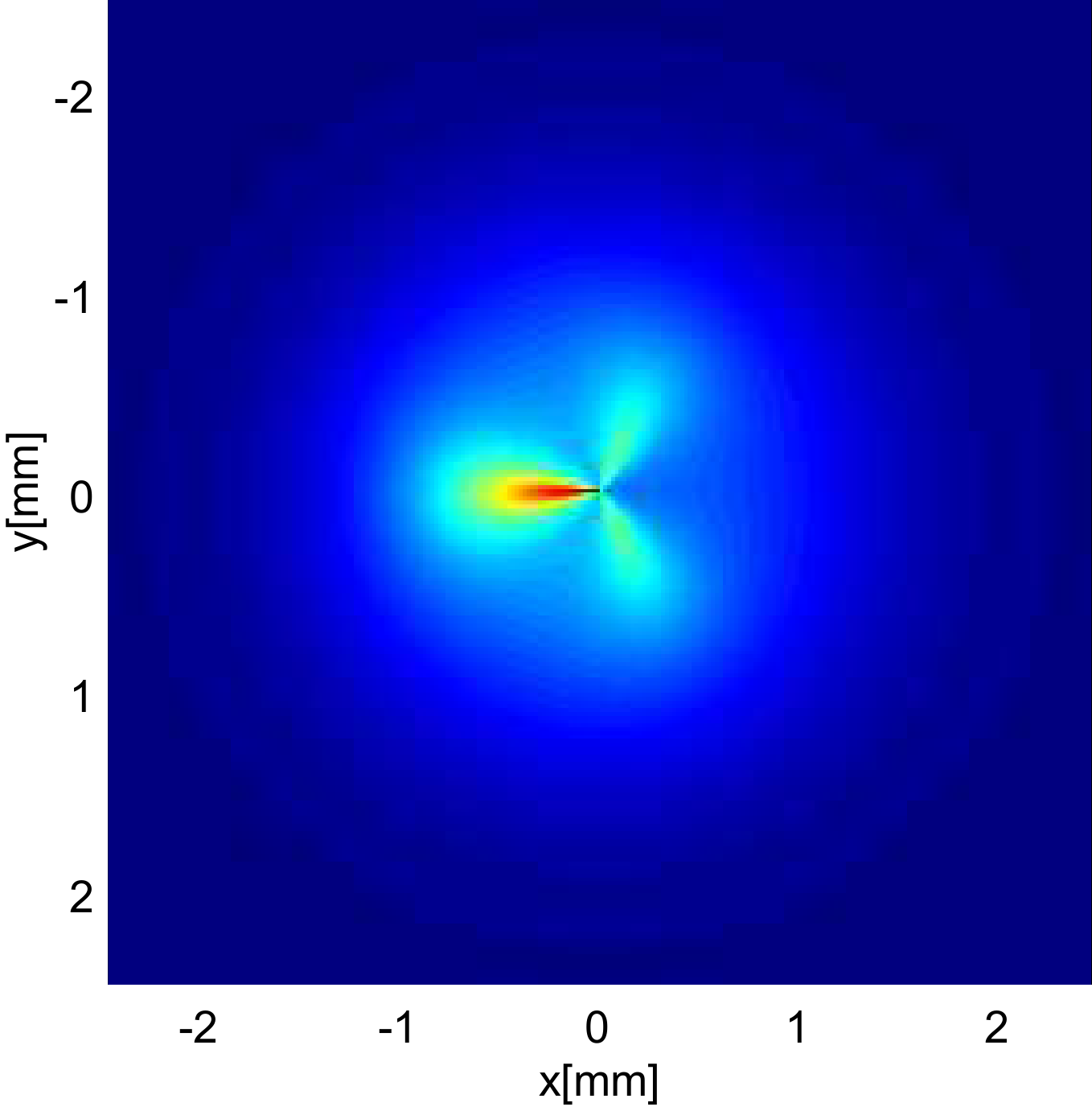}
\includegraphics[width=.2\textwidth]{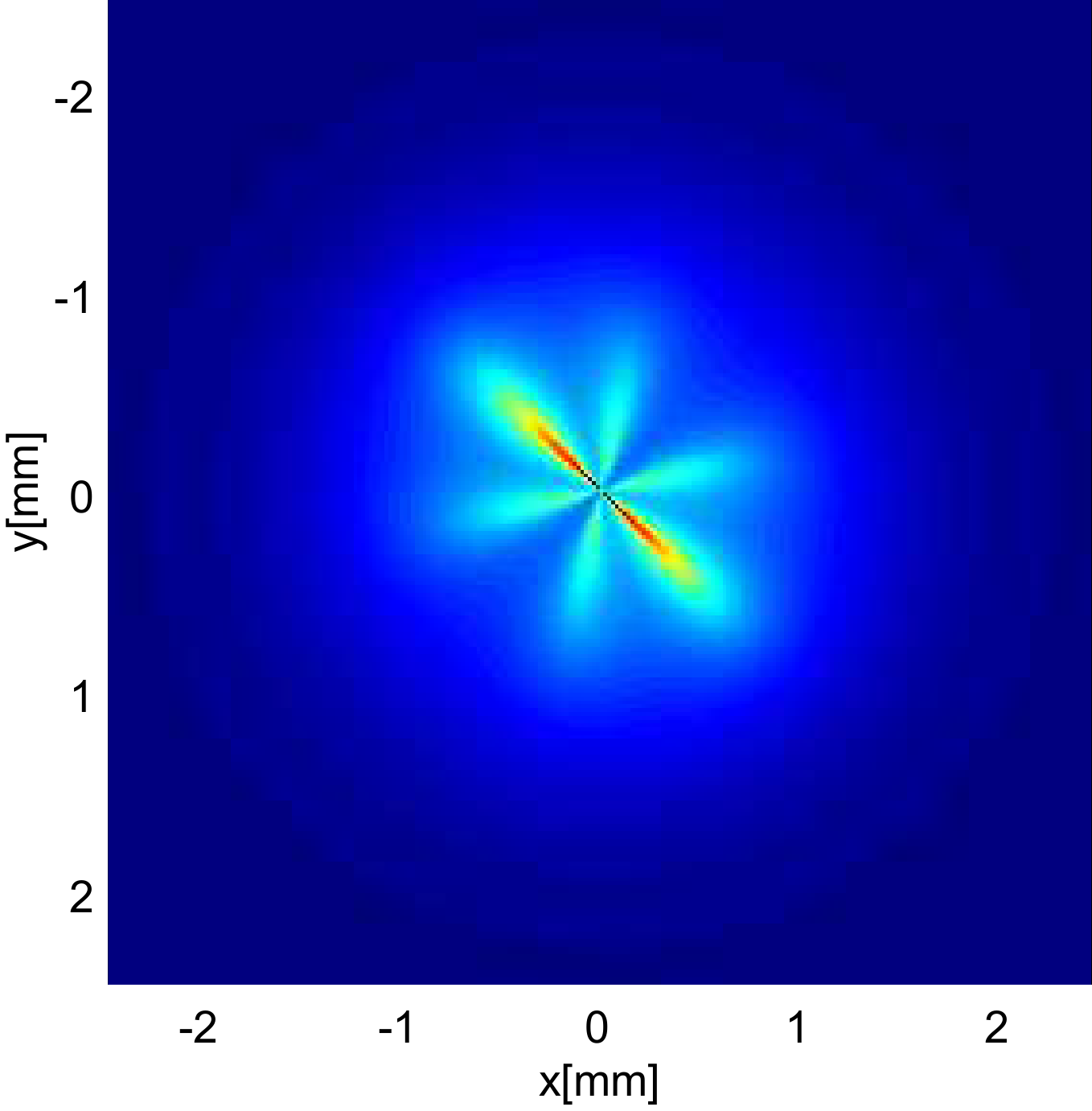}
\includegraphics[width=.2\textwidth]{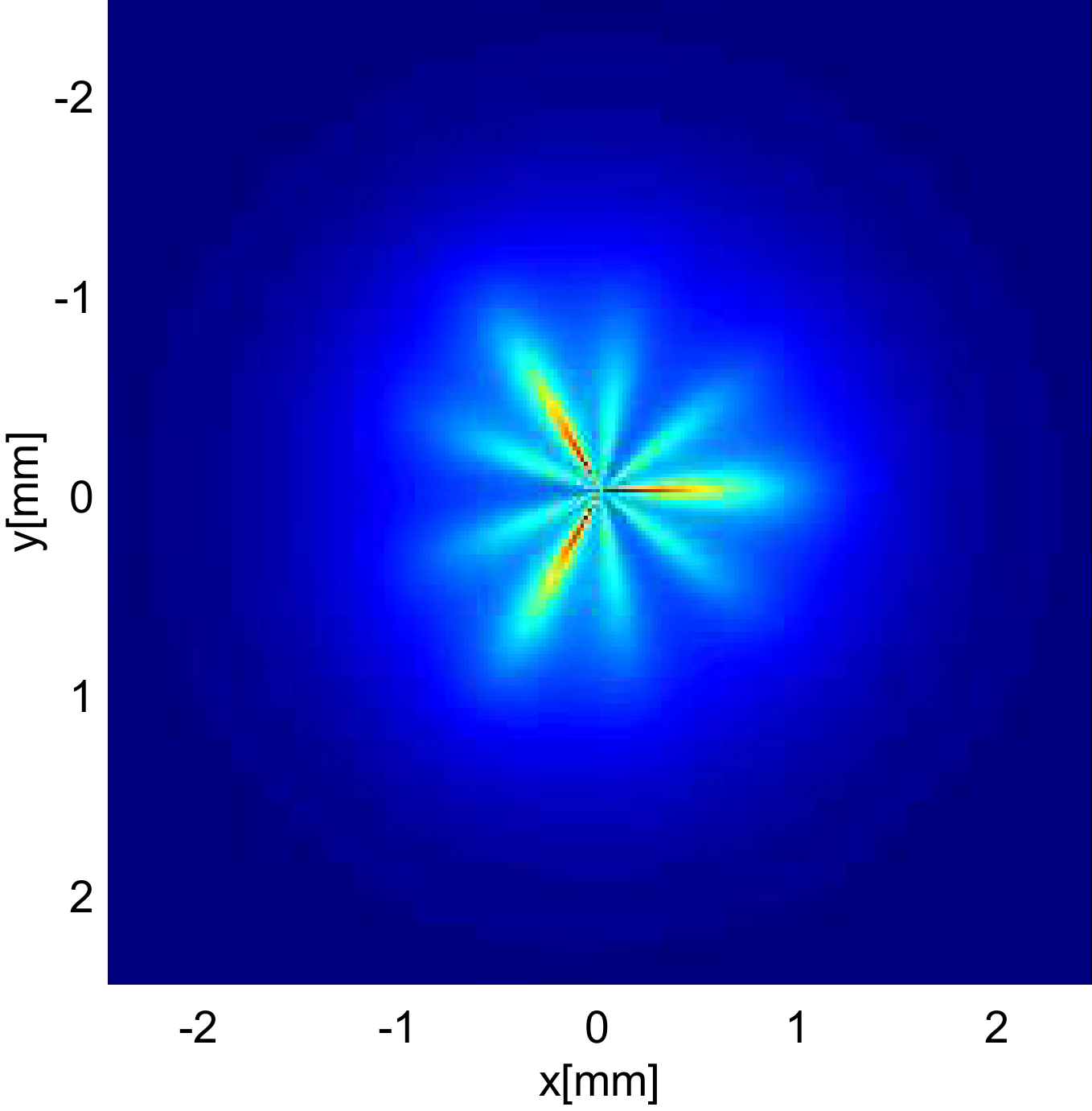}

(b)
\includegraphics[width=.2\textwidth]{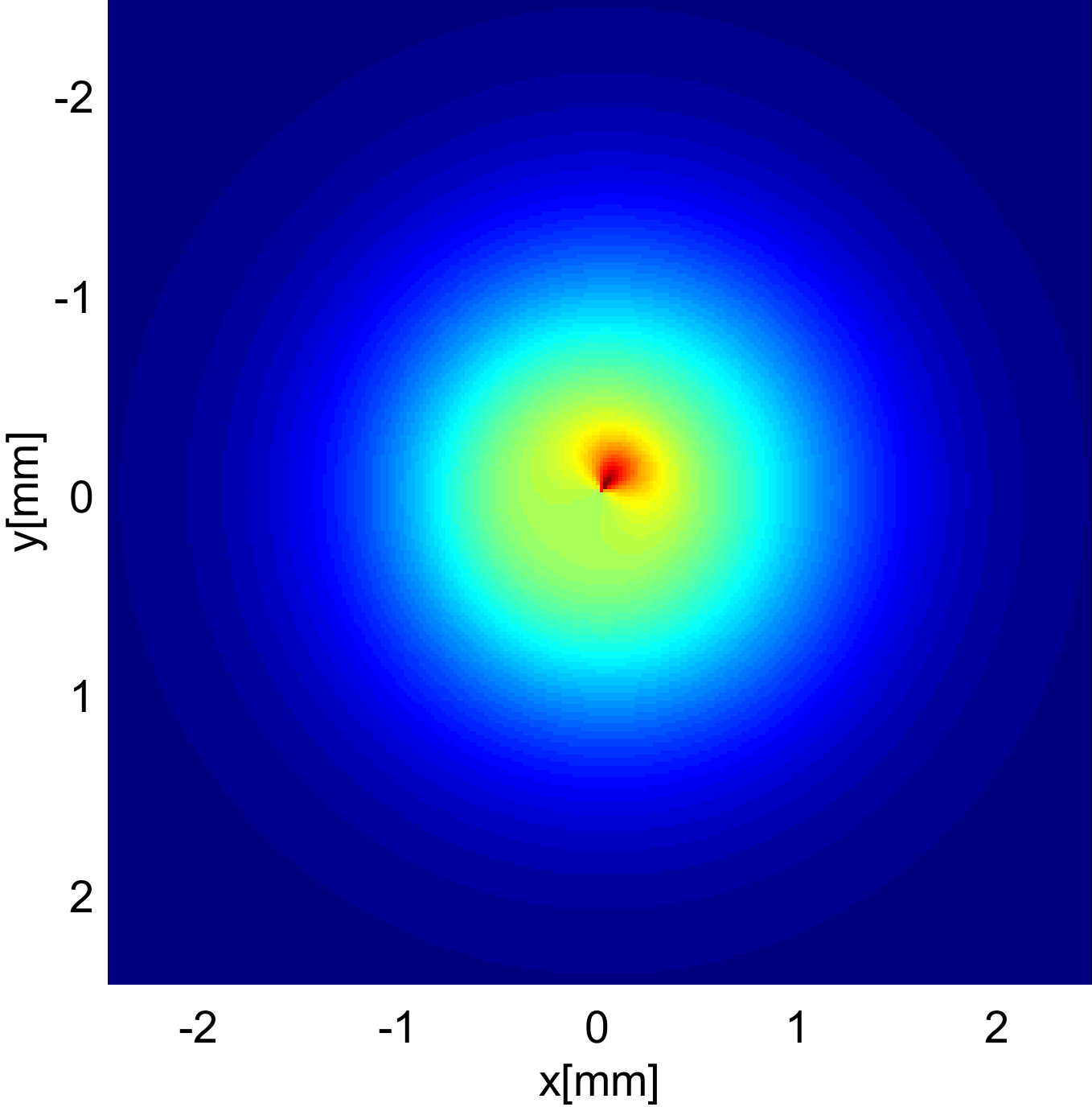}
\includegraphics[width=.2\textwidth]{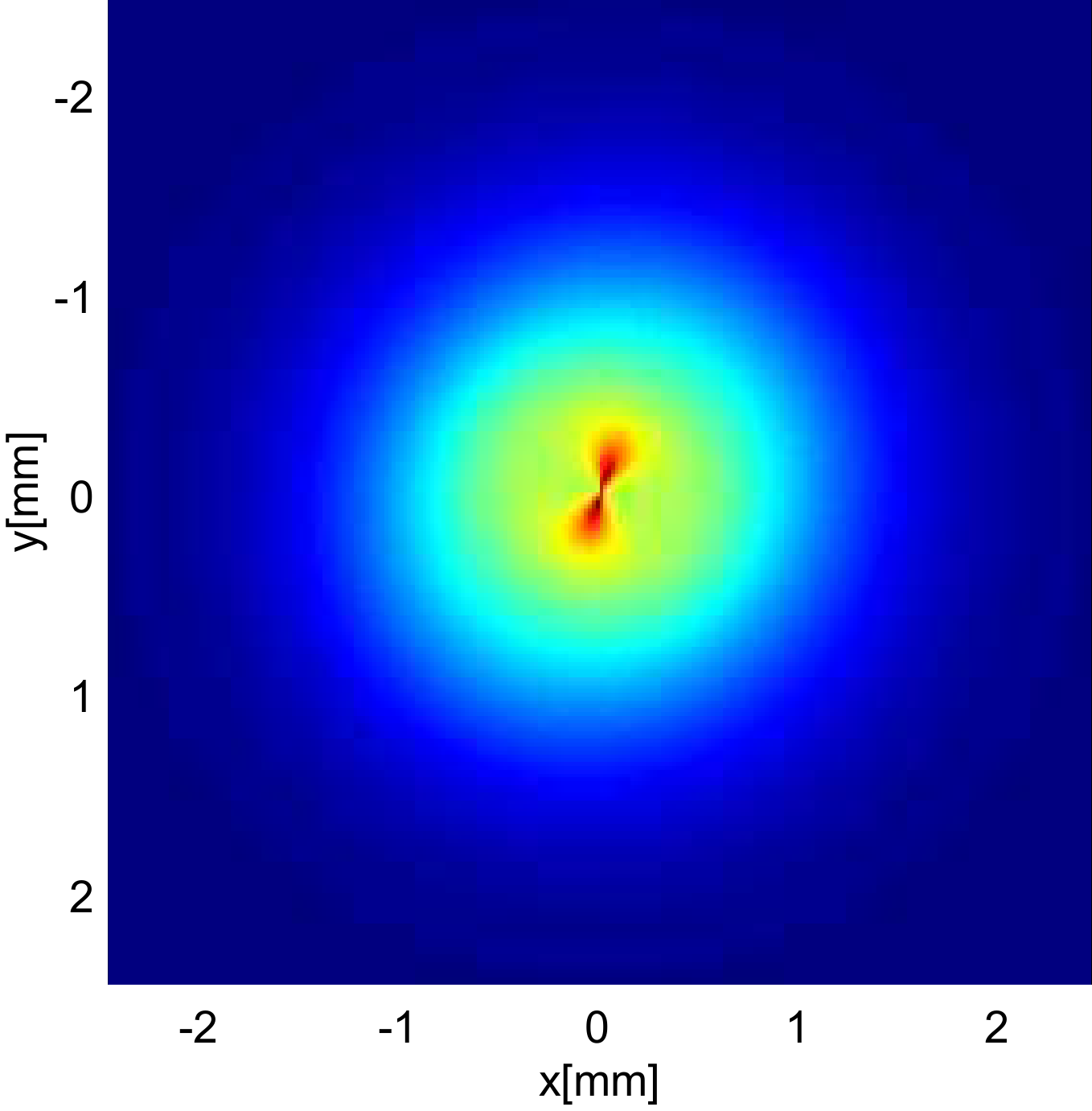}
\includegraphics[width=.2\textwidth]{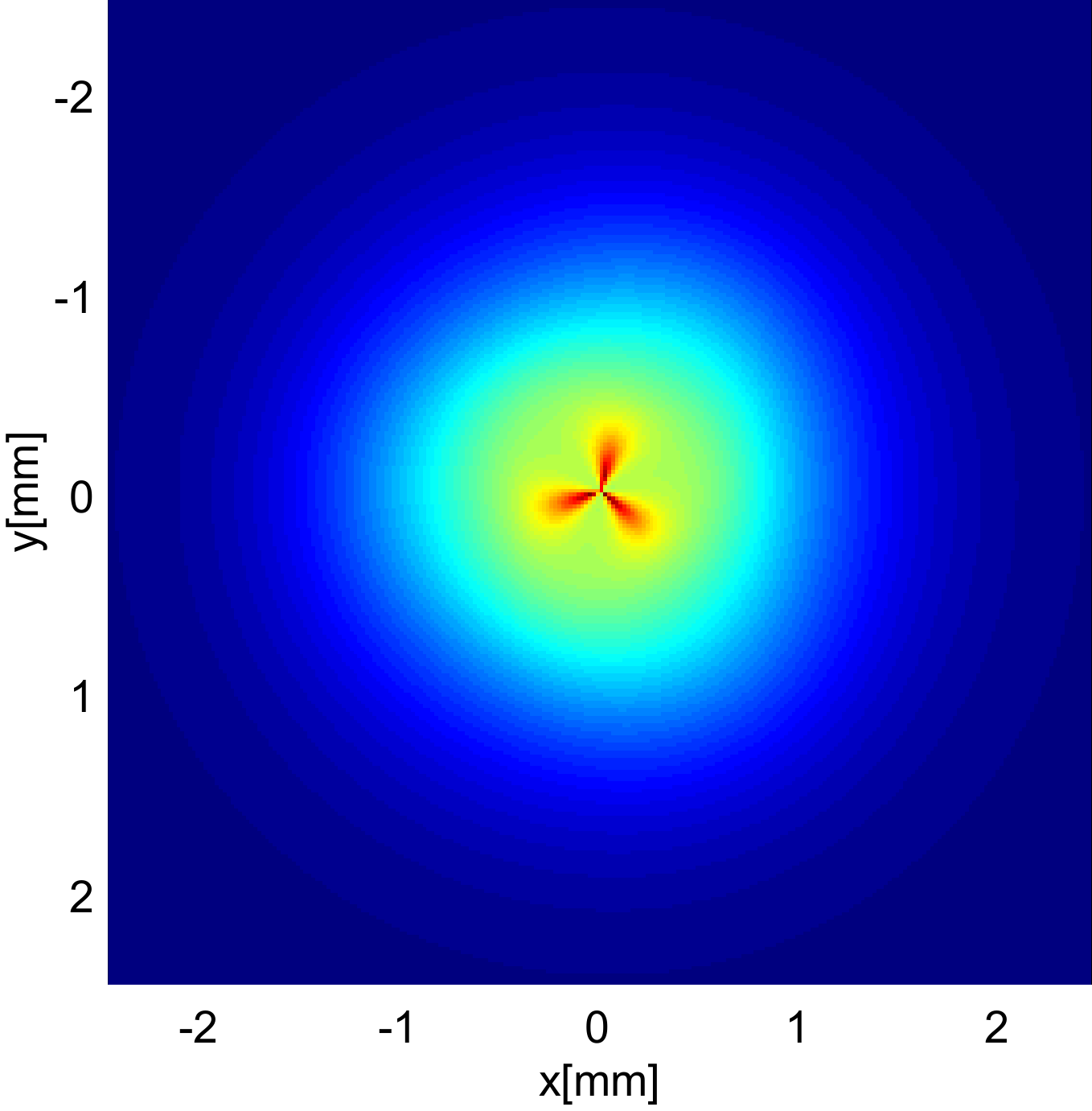}
\caption{The patterns for $Q=1,2,3$ vortices (left to right), in a dissipative (a) and lossless (b) left-handed metamaterial. The behavior for three different charges confirms the previous conclusions for the type of symmetry encountered. All  parameters except for the vortex charge are the same as for Fig.~\ref{pattvoc}. The propagation distance is $z=5b$ in (a) and $z=8b$ in (b).}
\label{pattvoc2}
\end{figure}

The findings above are further corroborated by Fig.~\ref{pattvoc2} which shows the vortices with different charges $Q=1,2,3$ in the same regime (dissipative left-handed, (a), and lossless left-handed, (b)). As claimed above, the
symmetry is $C_{3Q}$ in the panel (a), and (except at small $z$ values) $C_Q$ in the panel (b). Finally, it is obvious that there is some mixing of patterns: the polygons are never exactly regular, so the groups $C_n$ are certainly
not exact symmetries; we use the $C_n$-nomenclature merely for convenience.

One interesting phenomenon in Fig.~\ref{pattvoc}(c) is that the pattern rotates along the $z$ axis. This can be understood as excitation of multiple angular modes (of the form $e^{\imath lz}$ with various $l$-numbers) as the beam
travels along the sample. This is a well-known consequence of nonlinear terms \cite{rmp,book} and typically depends on the relative strength of nonlinear mode interactions compared to energy density $\vert E\vert^2+\vert H\vert^2$
and dissipation $\gamma$. We will not explore it in quantitative detail in this paper as it is only tangential to our main topic of radial symmetry breaking; as one can see, the structure remains the same; just the orientation changes.

One might rightly worry that the initial conditions which contain a vortex in both electric and magnetic field are not very realistic, as in most materials the electric field dominates the optical response.\footnote{We thank an
anonimous referee for pointing out this issue.} Therefore, in experimental practice, one typically prepares a vortex in the electric field making use of phase masks or some other method, and the initial magnetic field distribution is
completely analytic. In Appendix \ref{appb} we repeat the calculations from Fig.~\ref{pattvoc2} and show that the outcome is the same, including the vocabulary of patterns and their $C_n$ shapes. Therefore, the
$E$-$H$ symmetric ansatz is merely a matter of convenience, and the realistic regime where $\vert H\vert\ll\vert E\vert$ is in fact covered by our work.

Fig.~\ref{int-decay} shows that at long times the decay of intensity is universal for given dielectric dampening coefficient $\gamma$, which suggests the main mechanism of dissipation is in fact the radiative loss. This is because we
deliberately chose $\epsilon,\mu$ with small imaginary parts (for $\epsilon$ it can also be zero), so the losses in the medium are not so important when it comes to total energy (they are still important for being nonlinear and
influencing the patterns). One important difference between the lossless medium (black and blue symbols in Fig.~\ref{int-decay}) and the dissipative medium (red, magenta) is that the former has a short interval of growing intensity,
before reaching the universal regime of radiative decay. The physical reason is that the polarization, i.e., the rearrangement of charges in the self-defocusing metamaterial reduces the overall electrostatic potential energy of the
medium, and this energy becomes available to the beam, increasing its intensity. Clearly, once the radiative losses overcome the total potential energy available, the intensity decays. The growth is clearly a transient effect
which cannot persist for long $z$-intervals. A formal way to understand this is that the nonconservation of energy is encoded by the last term in (\ref{eome}), which can have a positive or negative imaginary part depending on the
sign of $\partial_z\mu/\mu$. At large values of $z$, we expect to enter a universal regime where this sign is constant, because the radiation loss dominates over nonlinearities and the exchange of energy between the beam and the
medium; this is the universal decay regime in the figure.

\section{The theory of vortex evolution}

The phenomenology described in the previous section can be understood on several levels. At the crudest level, we can introduce a variable-separation ansatz in the equations of motion and then linearize them in the amplitude (but not in the phase!). This picture explains the $C_{2Q}$ patterns, but not the $C_{3Q}$ and $C_Q$ regimes. It also does not explain the instabilities, that is the changes and disappearance of patterns during the $z$-propagation. For the full picture it is necessary to take into account the nonlinear effects through the loop corrections, i.e., to move perturbatively beyond the amplitude-linearized solution. A qualitative insight of the symmetry breaking can however be obtained also in a simpler and more elegant way, directly from the symmetry analysis of the model Lagrangian (\ref{lag}). Therefore, after finishing the amplitude-linearized analysis and the loop corrections from nonlinearity, we will obtain the same results from a unified mean-field treatment of the (nonlinear) model Lagrangian.

A note on terminology is in order. The solutions we find are not the textbook type of vortex with phase dependence solely of the type $e^{\imath Q\phi}$; rather, the dependence on the phase is more complicated, i.e., the phase is doing more than just the winding, but it is still true that the circulation of the phase around some point (the location of the vortex core) is an integer -- the topological charge of the vortex. Such solutions are sometimes called spirals \cite{rabin} whereas the term vortex is reserved for the simple winding-phase solutions. We nevertheless stick to the widespread term ''vortex'' for any topologically charged solution under the fundamental group of the $U(1)$ phase symmetry.

\subsection{Amplitude-linearized solution}

We will separate variables in the equations of motion (\ref{eome}-\ref{eomh}) (or the Lagrangian equations (\ref{eomlae}-\ref{eomlah}), which do not differ from the original equations at the amplitude-linearized level) and then plug in the vortex ansatz. The vortex ansatz is a solution which has a winding phase $\Phi$ with some winding number $Q$, for a constant (averaged) value of the permittivity $\mu_c=\mathrm{const.}$, because we ignore the nonlinear dependence of $\mu$ on $\vert H\vert$. The vortex solution of winding number (topological charge) $Q$ in cylindrical coordinates $(r,\phi,z)$ can be separated into regular and vortex parts:
\be
\label{vortsol0}E=E_\mathrm{reg}+E_\mathrm{vort},
\ee
We represent the vortex part as
\be
\label{ansatz}E_\mathrm{vort}(r,\phi,z)=Z_E(z)R_E(r)e^{\imath Q\phi-\imath\Phi(\phi)},
\ee
and analogously for the magnetic field. Along the $z$-axis we get $Z_E(z)=e^{\imath\lambda z}$ as expected, and the eigenvalue $\lambda$ is arbitrary for now, i.e., it is determined by the boundary conditions along the $z$-axis. Upon inserting (\ref{ansatz}) into (\ref{eome}), the equation separates into the angular part and the radial part. The former reads
\be
\label{vorteq}\Phi^{\prime\prime}-\imath\left(\Phi^\prime\right)^2+2\imath Q\Phi^\prime+\imath l^2=0,
\ee
where $l$ is the eigenvalue of the angular part. This is the crucial equation -- the phase dynamics is nonlinear because $\mu$ is in general complex and the terms with $\nabla_\perp\mu$ contain nonlinear dependence on the phase. The equation is easily solved by first introducing $w\equiv\Phi^\prime$ and then reducing it to quadratures. The outcome is
\be
\label{vortsol}\Phi(\phi)=\cos\left(\sqrt{Q^2+l^2}\phi+C_l\right).
\ee
In other words, we still stay with a winding solution but various winding numbers (equal to $\sqrt{Q^2+l^2}$) are possible when multiple modes are excited. Clearly, only the solutions with integer windings are physical, otherwise they would not be single-valued. The most general solution is thus a superposition of solutions $Z_E(\lambda,l;z)\Phi_E(\lambda,l;\phi)R_E(\lambda,l;r)$ with different $l$-modes so as to result in a single-valued function. Now the radial part acquires the form
\be
\label{prevorteqr}R_E^{\prime\prime}+\frac{1}{r}R_E^\prime+\left(\frac{\lambda}{r^2}-k^2+\epsilon_{D0}\tilde{\omega}^2\right)R_E+\frac{\alpha\mu_c\tilde{\omega}^2}{E_c^2}R_E^3=0,
\ee
with $\tilde{\omega}\equiv\omega (1-\omega_0^2/(\omega^2+\imath\omega\gamma))$. If we disregard the cubic term (amplitude-linearized approximation!),\footnote{This is justified at least in some interval of $z$ values, as the system is dissipative and loses power $\int (E^2+H^2)$, so the amplitude progressively decreases along $z$.} the well-known solution in terms of Bessel functions is obtained:
\bea
\nonumber R_E(r)&\approx &c^{(1)}_E(\lambda,l)J_{Q_l}\left(ar\right)+c^{(2)}_E(\lambda,l)Y_{Q_l}\left(ar\right)\\
\label{radsol}Q_l&\equiv&\sqrt{Q^2+l^2},~~a\equiv\sqrt{\lambda-\epsilon_{D0}\mu_c\tilde{\omega}/\omega E_0^2-k^2}.
\eea
Here, $J$ and $Y$ are the Bessel functions of first and second kind, respectively. Similar solutions $Z_H(z),\Phi(\phi),R_H(r)$ are obtained for the magnetic field. The angular equation is identical for both fields: for this reason
we have one solution $\Phi$ for both $E$ and $H$. The eigenvalues $\lambda,l$ and the values of the constants $c^{(1,2)}_{E,H}$ are determined by the boundary conditions. Obviously, (\ref{vortsol}) imposes the $C_{2Q}$ symmetry,
\emph{if} $l=0$. This simplest case is not necessarily the stable solution. We might have a sum over many $l$-values. In principle, such sums may yield more complicated patterns, however we will see that when the physically
reasonable boundary conditions are implemented (decay at infinity, single-valuedness everywhere) one typically always has the robust $C_{2Q}$ pattern. One important consequence of the fact that multiple $l$-modes are possible is that
due to nonlinear effects a new $l$-mode can be created during the propagation along the $z$-axis. We have already seen an example in Fig.~\ref{pattvoc}(c). A quantitative analysis of this phenomenon requires a full nonlinear model
and so can only be studied within the formalism of the next section.

\begin{figure}[H]\centering
(a)\includegraphics[width=40mm]{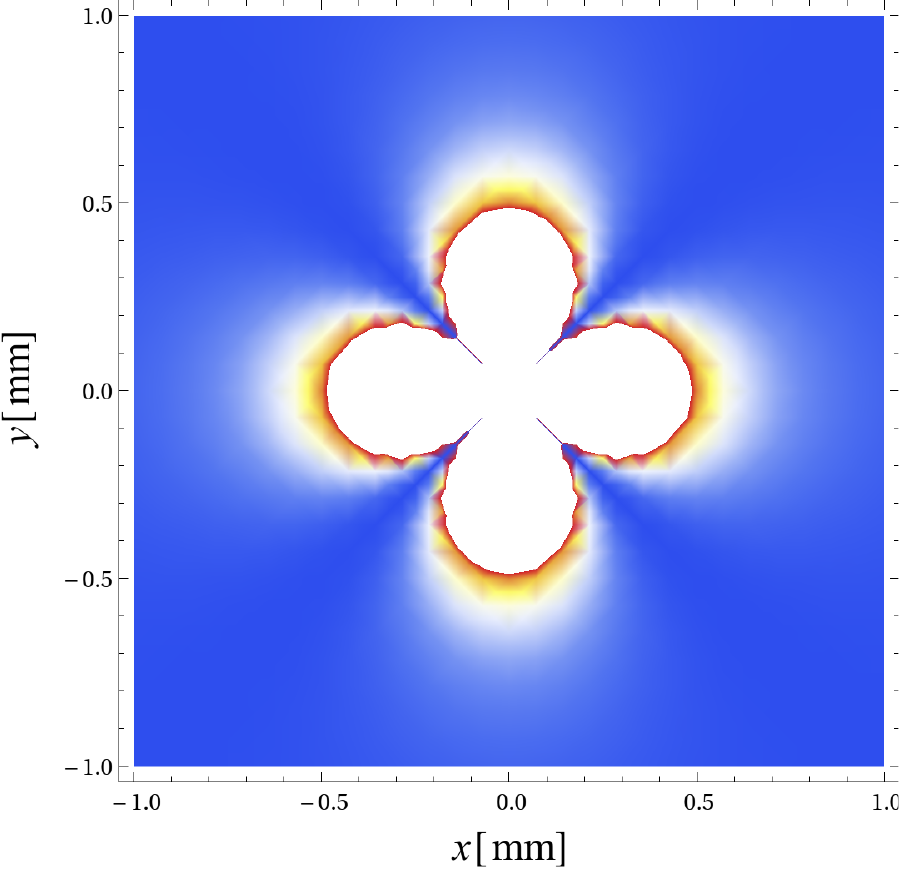}
(b)\includegraphics[width=40mm]{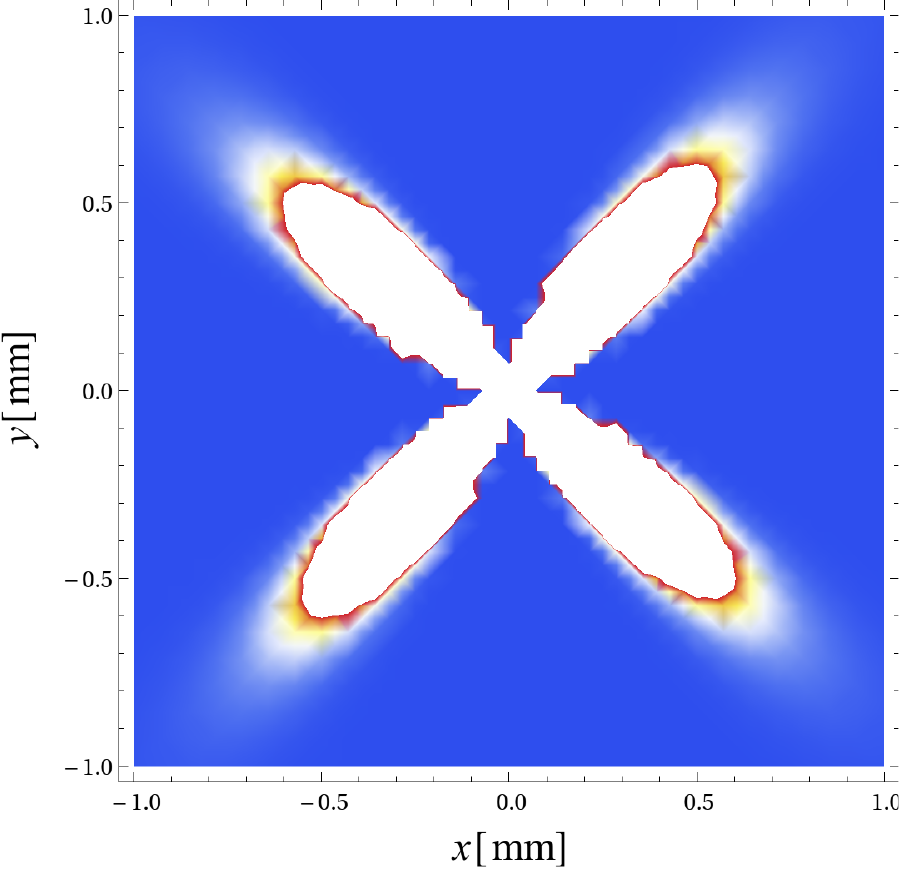}
\caption{\label{figpattern} Polygonal pattern $\vert E\vert^2$ for a vortex of charge $Q=2$, for $k=2,\epsilon_{D0}=12.8,\mu_c=1.004$ (values of all parameters and constants in the main text), at radial slice $z=1$, for a single
vortex mode $l=0$ (a), and for a linear combination of modes with $l=0,1,2$ decaying at infinity (b). The symmetry is $C_{2Q}=C_4$, which does not explain the $C_Q$ and $C_{3Q}$ regimes. Obviously, the crude picture of breaking the
radial symmetry works but full explanation is lacking. It will come from the loop corrections.}
\end{figure}

This solution is not very satisfying but reproduces some of the features from the numerics, summarized at the start of the previous section: (1) the reduction of the full $O(2)$ symmetry down to a discrete symmetry $C_n$ for some
$n\in\mathbb{N}$, i.e., the polygonal form of the vortex (2) the value $n=2Q$ is true in some but not in all situations. We show the solutions for a single angular mode from (\ref{vortsol},\ref{radsol}) in Fig.~\ref{figpattern}(a). In
Fig.~\ref{figpattern}(b), we show a linear combination of angular modes with $l=0,1,2$, with the coefficients $c^{(1,2)}_{E,H}$ in (\ref{radsol}) chosen so that the total intensity still decays sufficiently fast at infinity. The
symmetry is still $C_{2Q}$. Apparently, the regimes with the $C_Q$ and $C_{3Q}$ symmetries require loop corrections from nonlinear $\mu$ to be taken into account.

\subsection{Loop corrections}

The origin of the breaking of radial symmetry is the fact that a discrete set of modes in Fourier space is selected. This is best seen from the Fourier transform of the solutions (\ref{vortsol}) and (\ref{radsol}). We will calculate the propagator $G(\mathbf{u})$ at constant $z$, i.e., the Fourier transform $\mathbf{r}\mapsto\mathbf{u}$ of the solution with a Dirac delta source. This source imposes the boundary condition $R_E(0)\to\infty,\int drr\cos\phi R_E(r)=1$, giving $c^{(1)}_E=0,c^{(2)}_E=2\pi/\Gamma(Q/2)$ in (\ref{radsol}). Fourier-transforming $(x,y)\mapsto (u_x,u_y)$ we get for a single mode (\ref{radsol}), making use of the Bessel and Lommel integrals:
\bea
\nonumber G_{E,H}(\mathbf{u})=\frac{2\pi}{\Gamma(Q/2)}\frac{e^{\imath Q(\pi/2+\phi)}}{au}\left(\frac{\sin\left(\left(u-a\right)\Lambda\right)}{u-a}-\frac{\cos\left(\left(u+a\right)\Lambda-\pi Q\right)}{u+a}\right)+\\
\label{prop}+\frac{2\pi}{\Gamma(Q/2)}\frac{e^{-\imath Q(\pi/2+\phi)}}{au}\left(\frac{\cos\left(\left(u-a\right)\Lambda+\pi Q\right)}{u-a}-\frac{\cos\left(\left(u+a\right)\Lambda\right)}{u+a}\right).
\eea
Here, $\Lambda$ is the ultraviolet (UV) small-length/high-momentum cutoff, i.e., the Fourier transform is performed by integrating $\int_{1/\Lambda}^\infty dr\int_0^{2\pi} d\phi$. The cutoff has a clear physical meaning: $1/\Lambda$ is the size of the vortex core (where the vortex ansatz stops working because the gradient of the field becomes too high). We clearly do not get anything new by just Fourier-transforming. The goal is to move beyond the amplitude-linearized approximation of the previous section by considering the effects of non-constant permittivity $\mu$ instead of constant (averaged) $\mu_c$. This calculation is essentially elementary but might be tedious and boring for readers who are not fond of perturbative field theory. Most of the integrations are in Appendix \ref{appc}. Even the rest of this subsection can be skipped until the the equation (\ref{loopsol}) where we discuss the final result.

Putting $\mu$ from (\ref{om0eq1}) in place of $\mu_c$ requires the solutions for $\omega_{0NL}$ in terms of the magnetic field. The solutions are readily found from the Cardan formulas (we do not give them explicitly as they are cumbersome and not very illustrative). But the form of the $H$-dependence of $\omega_{0NL}$ is seen already from the Viete formula:
\be
\left(\omega_{0NL}^{(1)}\right)^2+\left(\omega_{0NL}^{(2)}\right)^2+\left(\omega_{0NL}^{(3)}\right)^2=\frac{1+2\Omega^2}{1+\vert H\vert^2/\alpha E_c^2},
\ee
so the solutions depend on $\vert H\vert^2$ only, with no higher powers of the magnetic field. Inserting this into $\mathcal{L}$, we get the nonlinear correction of the form:
\be
\label{muloop}\delta\mathcal{L}=g_{2,0,0}\vert\nabla_\perp E\vert^2+g_{0,2,0}\vert E\vert^2+g_{0,2,2}\vert E\vert^2\vert H\vert^2+g_{2,0,2}\vert\nabla_\perp E\vert^2\vert H\vert^2.
\ee
We thus have two quartic interaction terms and two quadratic terms. We do not intend to calculate the loop corrections in full detail; it is not worth the effort as we only want to capture the symmetry, i.e., the form of the angular dependence. First of all, the quadratic corrections $g_{2,0,0},g_{0,2,0}$ trivially renormalize the parameters in the bare propagator and do not change its functional form. Lowest-order nontrivial loop corrections to the self-energy come from $g_{0,2,2}$ and $g_{2,0,2}$. The electric field receives the correction $G_E^{-1}\mapsto\left(G_E+\Sigma_E^{(1)}+\Sigma_E^{(2)}\right)^{-1}$ with
\bea
\nonumber\Sigma_E^{(1)}&=&g_{0,2,2}\int d\mathbf{u}'G_H(\mathbf{u}')\approx g_{0,2,2}e^{3\imath Q/2}\sin(\pi Q)\log\Lambda\\
\label{loopcorrel}\Sigma_E^{(2)}&=&\frac{3}{2}g_{0,2,2}\int d\mathbf{u}'\int d\mathbf{\mathbf{u}}''G_H(\mathbf{u}')G_H(\mathbf{u}'')G_E(\mathbf{u}-\mathbf{u}'-\mathbf{u}'')\approx
\mathrm{const.}\times\left(a^{3/2}\cos(3Q\phi/2)-2\imath Q^2\log a\right).
\eea
We will write all equations for $E$, because this field receives interesting corrections from the gradient of $\mu$ in Eqs. (\ref{eome},\ref{eomlae}). The magnetic field does not couple to the permeability $\epsilon$ in the same way in the original equation (\ref{eomh}), and in the Lagrangian form (\ref{eomlah}) it does but $\epsilon$ does not contain such strong (non-polynomial) nonlinearities as $\mu$. One- and two-loop corrections appear not only in the self-energy but also in the vertex operators. However, the vertex corrections only have a weak momentum dependence and consequently the coordinate dependence (geometric patterns) of the solution is not significantly affected by them. For that reason we will not discuss them in detail.

The correction $\Sigma_E^{(1)}$ is the Hartree correction with a single vacuum bubble which is not very interesting: it merely introduces an additional mass term and does not influence the momentum dependence and thus the geometry of the patterns. As could be expected from power counting, it is logarithmically divergent in the UV cutoff $\Lambda$. Of course, this is not a problem in an effective theory; we have already explained the physical meaning of $\Lambda$. The watermelon diagram $\Sigma_{E,H}^{(2)}$ is crucial: it is momentum-dependent. Its calculation is found in Appendix \ref{appc}. An informal way to estimate its effect is the following: the leading contribution comes from the region where $\mathbf{u}\approx\mathbf{u}'-\mathbf{u}''$ because this is
a pole of the self-energy correction. Then we are left with angular integrals only, and they reduce to integrals of products of three rational functions (for the three propagators in (\ref{loopcorrel})) of the half-angle -- this gives rise to $3\phi/2$ in the argument of the cosine. Now the dressed propagator $\left(G_{E,H}^{-1}+\Sigma\right)^{-1}$ needs to be Fourier-transformed back to real space. We will only do this approximately (it is likely impossible to do exactly in closed form). The outcome is
\bea
\nonumber E_\mathrm{vort}(r,\phi,z)&=&\frac{e^{\left(\imath\lambda-2Q^2\log a\right)z}\cos(Q\phi)}{\sqrt{\kappa r}}\times\\
\label{loopsol}&\times&\left(c^{(1)}_E(\lambda,l)\left(1+\frac{(2\pi)^{3/2}g_{0,2,2}}{\Gamma(Q/2)^3}\cos\left(3Q\phi/2\right)\right)+
{c^{(2)}_E(\lambda,l)\left(1+\frac{(2\pi)^{3/2}g_{0,2,2}}{\Gamma(Q/2)^3}\sin\left(3Q\phi/2\right)\right)}\right).
\eea
No doubt the reader sees that the terms $\cos(3Q\phi/2),\sin(3Q\phi/2)$ give a pattern $\vert E_\mathrm{vort}\vert^2$ with $3Q$ branches, in addition to the $2Q$-polygons obtained from the term $\cos(Q\phi)$. The interference between the two patterns might (1) break the symmetry completely (2) lead to $C_Q$ symmetry if the relative phase between the leading term and the corrections is approximately $2\pi/Q$. Both cases are seen in numerical work: $C_{3Q}$ appears in all left-handed materials (Fig.~\ref{pattvoc}(a,c)), and elements of $C_Q$ symmetry are present in almost all cases at long propagation distances $z$ (Fig.~\ref{pattvoc}(a,b,c),~Fig.~\ref{pattvoc2}).

The self-energy has an imaginary part (equivalently, the solution (\ref{loopsol}) exhibits exponential decay in $z$), meaning that these configurations are not stable - they are only seen up to some propagation distance $z$. The exact order (along $z$) and stability of each of the patterns depends on the details of the permeability $\epsilon$. One important and universal lesson is however that the decay rate (the real part of the exponent in (\ref{loopsol})) is proportional to $Q^2$, therefore the higher the value of $\vert Q\vert$, the faster it decays. This supports the general intuition that vortices with high winding numbers are not stable. But unlike the simplest case of the XY model or a superfluid where the stability only allows $Q=\pm 1$, we can in principle have arbitrarily high $Q$ as we have seen also in the numerics; their lifetimes are smaller and smaller as $Q$ grows, but still finite. The exponential decay itself is also confirmed by the numerics, as seen from Fig.~\ref{int-decay}. 

\subsection{Isotropy breaking -- the look from the action}

The basic mechanism leading to the symmetry breaking $O(2)\mapsto C_{3Q}\mapsto C_{2Q}\mapsto C_Q$ is seen already from the model Lagrangian (\ref{lag}). The symmetry breaking is essentially the consequence of the interplay of the
nonlinear-sigma-model form of the kinetic term and the complex nonlinearity of the magnetic permittivity $\mu$. Therefore, we can take a static approximation of the $z$-dynamics, ignoring the $z$-dependence; clearly, in that
framework we can only obtain the vocabulary of patterns, not the relative stability of $C_Q,C_{2Q},C_{3Q}$.\footnote{We could take the ansatz $e^{\imath\lambda z}$ instead; it would merely modify $k^2\mapsto k^2-\lambda$.} The
separation of variables remains a natural ansatz, and the vortex nature of the solution implies $E_\mathrm{vort}=E_0(r)e^{\imath\Theta(\phi)}$ with $\oint d\phi\Theta(\phi)=2\pi Q$ and analogously for the magnetic field. The
Lagrangian (\ref{lag}) then becomes:
\be
\label{fren}\mathcal{L}=
\frac{\left(E_0^\prime\right)^2+\frac{\left(\Theta^\prime\right)^2}{r^2}+k^2E_0^2}{\mu}+\frac{\left(H_0^\prime\right)^2+\frac{\left(\Theta^\prime\right)^2}{r^2}+k^2H_0^2}{\epsilon}.
\ee
The fact that $\mu$ contains $\omega_{0NL}^2(\vert H\vert^2)$, which is in turn the solution of the cubic equation, introduces a branch cut in $H$ because of the cubic roots. This is the simplest explanation of the origin of the
$C_{3Q}$ symmetry. More quantitatively, the story follows exactly the Landau-Ginzburg paradigm: while the initial Lagrangian only depends on $\vert E\vert^2$ and $\vert H\vert^2$ and thus preserves isotropy, the saddle-point
solution is given by the equation
\be
\label{laeq}\frac{\epsilon\left(\nabla_\perp^2-H\right)E-\epsilon^\prime\nabla_\perp E\cdot\nabla_\perp H}{\epsilon^2}+\frac{\mu^\prime}{\mu^2}\vert H\vert^{-1/3}=0,
\ee
where we have used that $\mu=\mu(\omega_{0NL}^2)$ and $\omega_{0NL}^2=\omega_{0NL}^2(\vert H\vert^{2/3},\vert H\vert^{4/3})$ (from the Cardan formulas). With the ansatz adopted above, the amplitude equation for $E_0(r)$ is the
nonlinear amplitude equation (\ref{prevorteqr}). The equation for the phase part $\Theta$ is more interesting. It reads
\be
\label{laeq2}\frac{\left(\Theta^\prime\right)^2\left(1-\frac{\epsilon^\prime}{\epsilon}\frac{E_0}{H_0}\right)-k}{\epsilon}+\frac{2\mu^\prime}{3\mu^2}\vert H\vert^{-1/3}=0.
\ee
The cubic root carries a branch cut, and the last term really evaluates to $2\mu^\prime/3\mu^2\times H_0^{-1/3}e^{-\imath\Theta/3+2n\pi\imath/3}$ with $n=-1,0,1$. The solution $\Theta_0$ which satisfies the phase winding condition
is obtained in implicit form as
\be
\label{laeqsol}\imath(\Theta_0+2\pi n/3)=K_n\log\left[\frac{k\left(1-\frac{\epsilon^\prime}{\epsilon}\frac{E_0}{H_0}\right)}{E_0^2+H_0^2}\sec^2\left(\frac{Q}{2}\phi\right)\right],
\ee
where $K_n$ is a constant determined by the amplitude solution and depending also on $n=-1,0,1$; its exact value is hard to find analytically as we do not know the solution to the amplitude equation in the nonlinear regime. But that
is not crucial for our general argument. The point is that the system can choose a solution with any of the values $n=-1,0,1$, i.e., even though the equations of motion (and the Lagrangian) are isotropic, the solution is not. Each
$n$-branch behaves as $\sim 1/\cos^2(Q\phi/2)$, only they are rotated by $\pm 2\pi/3$ with respect to each other; and each of them has a $C_Q$ symmetry. Put together, the three branches give $C_{3Q}$ patterns. But all that holds if 
two of the cubic roots are complex. If all cubic roots are real, the phase remains single-valued, and we only have $C_Q$ symmetry, coming directly from (\ref{laeqsol}) if we fix $n=0$, i.e., if we only keep a single
branch.\footnote{We use the fact that a cubic equation has either one or all three solutions real.}

What is the regime in which cubic roots are real and the symmetry is $C_Q$, as opposed to the complex roots and $C_{3Q}$ patterns? The easiest way is to look at the cubic equation (\ref{om0eq1}) for the magnetic permeability (and
the nonlinear frequency $\omega_{0NL}$). For $\mu>0$ (right-handed regime), the roots are all real and $C_{3Q}$ patterns cannot occur. Indeed, the $C_{3Q}$ phase is only present in Fig.~\ref{pattvoc}(a,c), in left-handed media.

%Starting from the mean-field solution (\ref{laeqsol}), we can expand the Lagrangian $\mathcal{L}\mapsto\mathcal{L}+\delta\mathcal{L}$ and obtain the fluctuation equation as
%\be
%\label{flucteq}\delta\Theta+\frac{}{\cos^2\Theta_n}e^{2\pi\imath n/3}\delta\Theta=\frac{1}{\cos^2\Theta}e^{2\pi\imath n/3}.
%\ee

This approach is much more physical and elegant than the tour-de-force calculations of the previous two sections but it does not give explicit solutions for $E,H$; it only classifies the symmetries of the solution. This is why we we
still needed the perturbative linear and two-loop analysis, to arrive at more quantitative results.

The saddle-point solution (\ref{laeqsol}) is nonlinear, unlike the linearized solution found in the first subsection (\ref{vortsol}). It is not a vacuum in the usual field-theory sense however, as it is not constant. We are dealing
with dynamical criticality of the kind discussed in \cite{cross}. In the vicinity of this solution, the Lagrangian describes the fluctuations of amplitude $\delta E,\delta H$, and the fluctations of phase $\delta\Phi$. Similar to
the $O(3)$-type spin models \cite{vortbook} and multi-beam optical systems \cite{nasstari}, and unlike simple XY-type models, the phase and amplitude fluctuations mix. By analyzing the fluctuation equations, it should be possible
to understand analytically also the transition from the left-handed to the right-handed regime as the parameters are varied, i.e., what are the instabilities that drive it. We will not attempt that here; it is a long subject that
deserves separate work.

\section{Toward experimental verification and applications}

We will now briefly discuss what an experimentalist can learn from our results and what to look for in practical work. Wave propagation through the metamaterial can be observed by measuring the transmission coefficients $S_{ij}$.
From these coefficients, one can also reconstruct the electric field intensity $\vert E\vert^2$, which can be directly compared to our intensity maps like Figs.~\ref{pattvoc}, \ref{pattvoc2} \cite{alitalo}. Another quantity which can
be measured is the voltage waveform, which can be used to construct amplitude envelopes \cite{kozyrev}.

Therefore, the predicted symmetry breaking is in principle directly observable. But the question remains how widespread it will be for realistic values of the parameters. From a more applied viewpoint, this question is reversed: how
to make a vortex transmission through a left-handed waveguide stable. In other words, how \emph{not} to observe the symmetry breaking. It is true that the phenomenon disappears as soon as the vortex charge is zero, i.e., when the
beam is not a vortex. However, the vortex patterns are likely important in applications. First, as a topologically protected object with conserved charge, a vortex is among the natural candidates for computational devices and
information transmission (for the same reasons that solitons are also interesting in that regard: they are robust to noise, carry a discrete "quantum" number, i.e., charge and are stable to small local perturbations). Second, in the
presence of impurities in the sample, vortices can form in a nonlinear metamaterial from the initially non-vortexing beam \cite{vortbook}.

\begin{figure}[H]\centering
(a)\includegraphics[width=80mm]{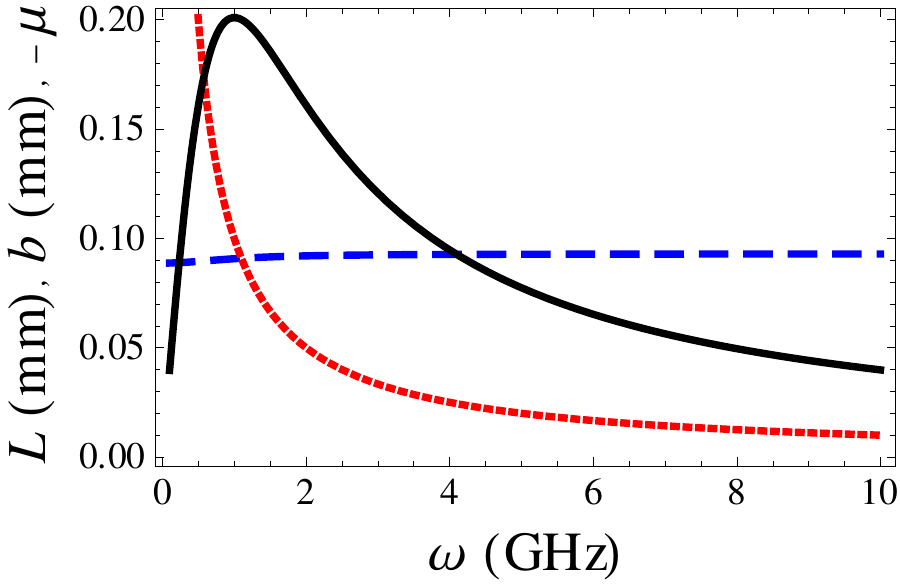}
(b)\includegraphics[width=80mm]{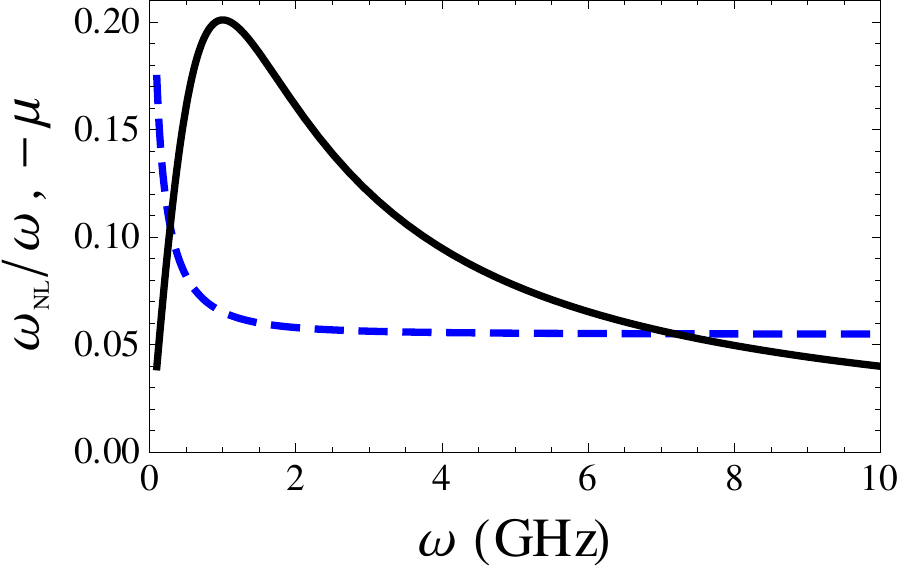}
\caption{\label{figtest} (a) Frequency dependence of the typical propagation length scale for the dissipation of the vortex $a^{2Q^2}$ (blue dashed line) and for the evolution of the symmetry-breaking $C_n$ patterns (red dotted
line). The symmetry breaking is detectable as long as the pattern evolution is faster than the dissipation, i.e., as long as the red curve is below the blue one. This is obviously the case for most of the frequency range. We also
plot the frequency dependence of the negative permittivity $-\mu$ (black full line; because of the minus sign large positive values in the plot are really large negative values of $\mu$). The left-handed regime is most prominent at
intermediate frequencies, which are also inside the regime of the symmetry breaking. (b) Frequency dependence of the relative strength of nonlinear interactions $\omega_{0NL}/\omega$ (blue dashed line) together with negative
permittivity $-\mu$ as in (a) (black full line). Our calculations, based on a pair of nonlinear Schr\"{o}dinger-like equations are reliable as long as the nonlinearity is not too strong. This is again the case for all but very
small frequencies, and again includes the left-handed regime.}
\end{figure}

Let us focus on the left-handed regime, which is the most interesting and the most relevant for applications. The first condition is therefore to be in the frequency regime with $\mu(\omega)<0$. This can be checked directly from
Eq.~(\ref{mueq}) as we did in Fig.~\ref{material}(c). The second issue is that the symmetry breaking takes some finite time, i.e., some finite propagation length, which is of order $b$; as can be seen from Fig.~\ref{pattvoc2} and
directly from Eqs.~(\ref{eome}-\ref{eomh}), this is the length scale over which the patterns change. On the other hand, the one-loop calculation (\ref{loopsol}) shows that the intensity decays with the rate $\sim a^{-2Q^2}$. As long
as this is less than the characteristic length $b$, one will likely not see the symmetry breaking but just eventual dissipation of the beam. Therefore, these two scales should be compared for some reasonable parameter values. We
show this in Fig.~\ref{figtest}(a) for $F=0.4$, $\epsilon_{D0}=12.8$, $\gamma=1$ GHz, $\omega_0=10$ GHz. Apparently, the length scale of the $C_n$ pattern development (red dotted line) is nearly always shorter than the dissipation
scale (blue dashed line), so we expect that the effect predicted in the paper is readily seen in experiment, at least for $Q=\pm 1$. For larger vortex charges, the dissipation grows quickly and high $Q$ values are probably not easily
observed. Conversely, if the goal is to keep a stable radially symmetric vortex pattern, one should remain at small frequencies, although for $\omega\ll\omega_0$ the material is not strongly left-handed, as can be seen from the
$-\mu(\omega)$ dependence, also given in the figure.

There is still one remaining issue. Our theoretical approach, based on a pair of nonlinear Schr\"{o}dinger-like equations, inherently disregards some effects. It describes a quasimonochromatic wave without wave mixing or dissipation
due to higher harmonic generation \cite{rmp}. Such phenomena become significant for strong nonlinearities, so we should compare the nonlinearities in $\epsilon$ and $\mu$ to the typical energy (frequency) scale of the vortex. In
Eqs.~(\ref{epseq}-\ref{mueq}) the approximate ratios of the nonlinear to linear terms are given by $\vert E\vert^2/\epsilon_{D0}$ and $\omega_{0NL}/\omega_0\sim (A/H)^{1/3}$. The first scale is frequency-independent and solely
depends on the beam intensity. The second scale depends on frequency and needs to be inspected more closely. In Fig.~\ref{figtest}(b) we plot the nonlinearity ratio for the magnetic field for a range of frequencies $\omega$, again
together with the permittivity to make sure we are at the same time in the left-handed regime. The relative nonlinearity strength quickly saturates around a value $0.06\ll 1$, so we are rather confident that our equations of motion
still make sense.

Altogether, the conclusion is that the breaking of radial symmetry is observable by standard means (measuring the transport coefficients and reconstructing the intensity map at the exit face of the metamaterial), as long as the 
frequency of the wave is not too low. This kind of instability kicks in at shorter propagation lengths (of order $0.1$ mm in Fig.~\ref{figtest}(a)) than the nonlinear diffraction effects studied for breathers in
\cite{boardman}, suggesting that vortex signals are more fragile and less convenient for information transmission.

\section{Discussion and conclusions}

Our main result is contained already in the title -- left-handedness and nonlinearity together create the breaking of the $O(2)$ symmetry down to a discrete group, with the pattern vocabulary consisting of the $C_{3Q},C_{2Q},C_Q$
patterns. In the right-handed system with the same nonlinearity the isotropy is broken again, but the pattern vocabulary only has $C_{2Q}$ and $C_Q$ stages. How exactly the patterns evolve into each other and through which
instabilities is not universal, and it depends on the exact form of $\epsilon$ and $\mu$. In our model, the $\epsilon$-dependence is mainly encapsulated in the dissipation $\gamma$: the left-handed non-dissipative case is usually
dominated by $C_Q$ after a much shorter $C_{3Q}$ phase, whereas the dissipative left-handed metamaterials most prominently show $C_{3Q}$ patterns. For the right-handed materials, non-dissipative and dissipative dynamics shows mainly
$C_{2Q}$ and $C_Q$ patterns, respectively.

A detailed account of the pattern dynamics was only possible through numerical work. But the vocabulary itself -- the existence of symmetries $C_{3Q},C_{2Q},C_Q$ -- we were able to understand analytically. The dynamic
Landau-Ginzburg picture reveals this as a consequence of the cubic root nonlinearity in the magnetic permittivity, \emph{and} the fact that the cubic equation has either two complex roots in the left-handed regime, or all three real
roots in the right-handed regime, \emph{and} the presence or absence of dissipation in the electric permeability. In the framework of our field theory model, the second derivative of the free energy (on-shell Lagrangian,
Landau-Ginzburg functional) likely has a jump when the symmetry changes. This is a strong encouragement that the phenomena we observe here, and in general the walk through the pattern vocabulary, can be understood from the viewpoint
of order/disorder transitions.

Similar phenomena were studied also in \cite{fan,kevrek} and above all \cite{walasik}, where $C_{3Q}$ necklaces were found, within a model of left-handed metamaterials given in \cite{fan} and
similar to ours. Clearly, we have not exhausted this subject; more research is still needed to fully understand the transition between different symmetries and their instabilities. Vortices in metamaterials seem to be a promising
arena, as in a metamaterial the nonlinearity and the frequency band where the materials is left-handed can to some extent be tuned at will. Therefore, the phase diagram of collective vortex interactions can also be studied, and is an
obvious topic for future work.

\section*{Acknowledgments}

This work has made use of the Sci-Hub service. We are grateful to Milan Petrovi\'{c} and Mariya Medvedyeva for helpful remarks. We also thank the referees for some important and stimulating questions. Work at the Institute of Physics is funded by Ministry of Education, Science and Technological Development, under grant no. OI171017.

\appendix

\section{Derivation of the equations of motion from the Maxwell equations}\label{appa}

Start from the definitions $\hat{D}=\epsilon\hat{E},\hat{B}=\mu\hat{H}$ and the Maxwell equations in absence of external charges and currents ($\rho=\hat{j}=0$):
\be
\nabla\cdot\hat{D}=\rho=0,~\nabla\cdot\hat{B}=0,~\nabla\times\hat{E}=-\partial_t\hat{B},~\nabla\times\hat{H}=4\pi\hat{j}+\partial_t\hat{D}=\partial_t\hat{D}.
\ee
We make the following assumptions:
\begin{enumerate}
 \item Small gradients of the permittivities $\epsilon$ and $\mu$, so their second and higher derivatives are disregarded. Since $\omega\propto k$, it means that mixed derivatives of the from $\partial_t\nabla \epsilon$ are also disregarded. In other words, the characteristic length scale $l$ along the $z$-axis on which $\epsilon,\mu$ change is assumed to be large compared to the characteristic scale $b$ of the changes in $E,H$.
 \item The time dependence is harmonic so $\partial_t=-\imath\omega$.
\end{enumerate}
Acting on the last equation by $\nabla\times$ and making use of the identity $\nabla\times\nabla\times\hat{H}=-\nabla^2\hat{H}+\nabla(\nabla\cdot\hat{H})$, one gets for the left-hand side:
\bea
\nonumber & &\nabla\times\nabla\times\hat{H}=-\nabla^2\hat{H}+\nabla\left(\nabla\cdot\frac{\hat{B}}{\mu}\right)=-\nabla^2\hat{H}+\nabla\left(\frac{1}{\mu}\nabla\cdot\hat{B}\right)-
\nabla\left(\frac{\nabla\mu}{\mu^2}\cdot\hat{B}\right)=\\& &=-\nabla^2\hat{H}+\nabla\left(\frac{1}{\mu}\nabla\cdot\hat{B}\right)-\nabla\cdot\left(\frac{\nabla\mu}{\mu^2}\right)\hat{B}-\frac{\nabla\mu}{\mu^2}\nabla\cdot\hat{B}=-\nabla^2\hat{H}+0+O(1/l^2)+0=-\nabla^2\hat{H},
\eea
where we used $\nabla\cdot\hat{B}=0$ and disregarded the second derivative of $\mu$. The right-hand side yields
\bea
\nonumber & &\nabla\times\nabla\times\hat{H}=\nabla\times(\partial_t\hat{D})=-\imath\omega\nabla\times\hat{D}=-\imath\omega\nabla\times(\epsilon\hat{E})=-\imath\omega(\nabla\epsilon)\hat{E}-
\imath\omega\epsilon\nabla\times\hat{E}=\\
& &=-\imath\omega(\nabla\epsilon)\hat{E}-\omega^2\epsilon\mu\hat{H}=O(1/l^2)+\omega^2\epsilon\mu\hat{H},
\eea
so we obtain
\be
\nabla^2\hat{H}+\omega^2\epsilon\mu\hat{H}=0.
\ee
For the $\hat{E}$ field we start from the third Maxwell equation, act by $\nabla\times$ and find for the left-hand side:
\bea
\nonumber & &\nabla\times\nabla\times\hat{E}=-\nabla^2\hat{E}+\nabla(\nabla\cdot\hat{E})=-\nabla^2\hat{E}-\nabla\left(\nabla\cdot\frac{\hat{D}}{\epsilon}\right)=\nabla^2\hat{E}-
\nabla\left(\frac{1}{\epsilon}\nabla\cdot\hat{D}\right)+\nabla\left(\frac{\nabla\epsilon}{\epsilon^2}\right)\epsilon\hat{E}+\\
& &+\frac{\nabla\epsilon}{\epsilon^2}\nabla\cdot\hat{D}=-\nabla^2\hat{E}+0+O(1/l^2)+0=-\nabla^2\hat{E},
\eea
and for the right-hand side we get
\bea
\nonumber & &\nabla\times\nabla\times\hat{E}=-\partial_t(\nabla\times\hat{B})=-\partial_t(\nabla\times\hat{B})=-\partial_t\left(\nabla\times\left(\mu\hat{H}\right)\right)=
-\partial_t\left(\left(\nabla\mu\right)\hat{H}+\mu\nabla\times\hat{H}\right)=
\\& &=-\left(\partial_t\nabla\mu\right)\hat{H}-\nabla\mu\cdot\partial_t\hat{H}-\partial_t\left(\mu\partial_t\hat{D}\right)=O(1/l^2)-\frac{\nabla\mu}{\mu}\nabla\hat{E}+\omega^2\epsilon\mu\hat{E},
\eea
so
\be
\nabla^2\hat{E}+\omega^2\epsilon\mu\hat{E}-\frac{\nabla\mu}{\mu}\nabla\hat{E}=0.
\ee
For our geometry we take the paraxial beam approximation, with the ansatz $\hat{E}=E(x,y)e^{\imath(kz-\omega t)},\hat{H}=H(x,y)e^{\imath(kz-\omega t)}$, so the nabla acts as
\be
\nabla\hat{E}=\left(\nabla_\perp E,\partial_zE+\imath kE\right)e^{\imath(kz-\omega t)},
\ee
and the Laplacian operator acts as
\be
\nabla^2\hat{E}=\left(\nabla_\perp^2E+2\imath k\partial_zE-k^2E\right)e^{\imath(kz-\omega t)},
\ee
and analogously for the magnetic field. Now to write the equations motion in the final from we rescale $E\to E\cdot 2kb,H\to H\cdot 2kb,z\mapsto z\cdot 2kb$, where $b$ is some characteristic length scale along the $z$-axis, and divide the equations by $bk^2$ to obtain the equations (\ref{eome}-\ref{eomh}), reprinted here for convenience:
\bea
\label{eomeapp}-\frac{\imath}{b}\partial_zE&=&\nabla_\perp^2E+\left[\omega^2\epsilon\left(\vert E\vert^2\right)\mu\left(\vert H\vert^2\right)-k^2\right]E-
\frac{\nabla_\perp\mu\left(\vert H\vert^2\right)}{\mu\left(\vert H\vert^2\right)}\nabla_\perp E-\imath\frac{\partial_z\mu\left(\vert H\vert^2\right)}{2\mu\left(\vert H\vert^2\right)}E\\
\label{eomhapp}-\frac{\imath}{b}\partial_zH&=&\nabla_\perp^2H+\left[\omega^2\epsilon\left(\vert E\vert^2\right)\mu\left(\vert H\vert^2\right)-k^2\right]H,
\eea
For comparison to the equations given in \cite{zarov03,sadriv03,sadriv05}, one needs (1) to rescale $H\mapsto\omega^2/c^2 H$ to get the term $-\gamma^2H=-k^2/\omega^2$ in (\ref{eomhapp}) and (2) to absorb the factor $-k^2$ in (\ref{eomeapp}) in the definition of $\epsilon_{D0}$. This is possible as $\epsilon$ and $\mu$ have a constant term (equal $\epsilon_{D0}$ and $1$ respectively) so the product $\epsilon\mu$ also has a constant term proportional to $\epsilon_{D0}$, and the contribution $k^2E$ can be absorbed as $\epsilon_{D0}\mapsto\epsilon_{D0}-k^2$. We thus arrive at a system identical to that from \cite{zarov03}, except for the extra terms for the propagation along the $z$-axis.

\section{Configurations with no vorticity in the magnetic field}\label{appb}

\begin{figure}[H]
\centering
(a)
\includegraphics[width=.31\textwidth]{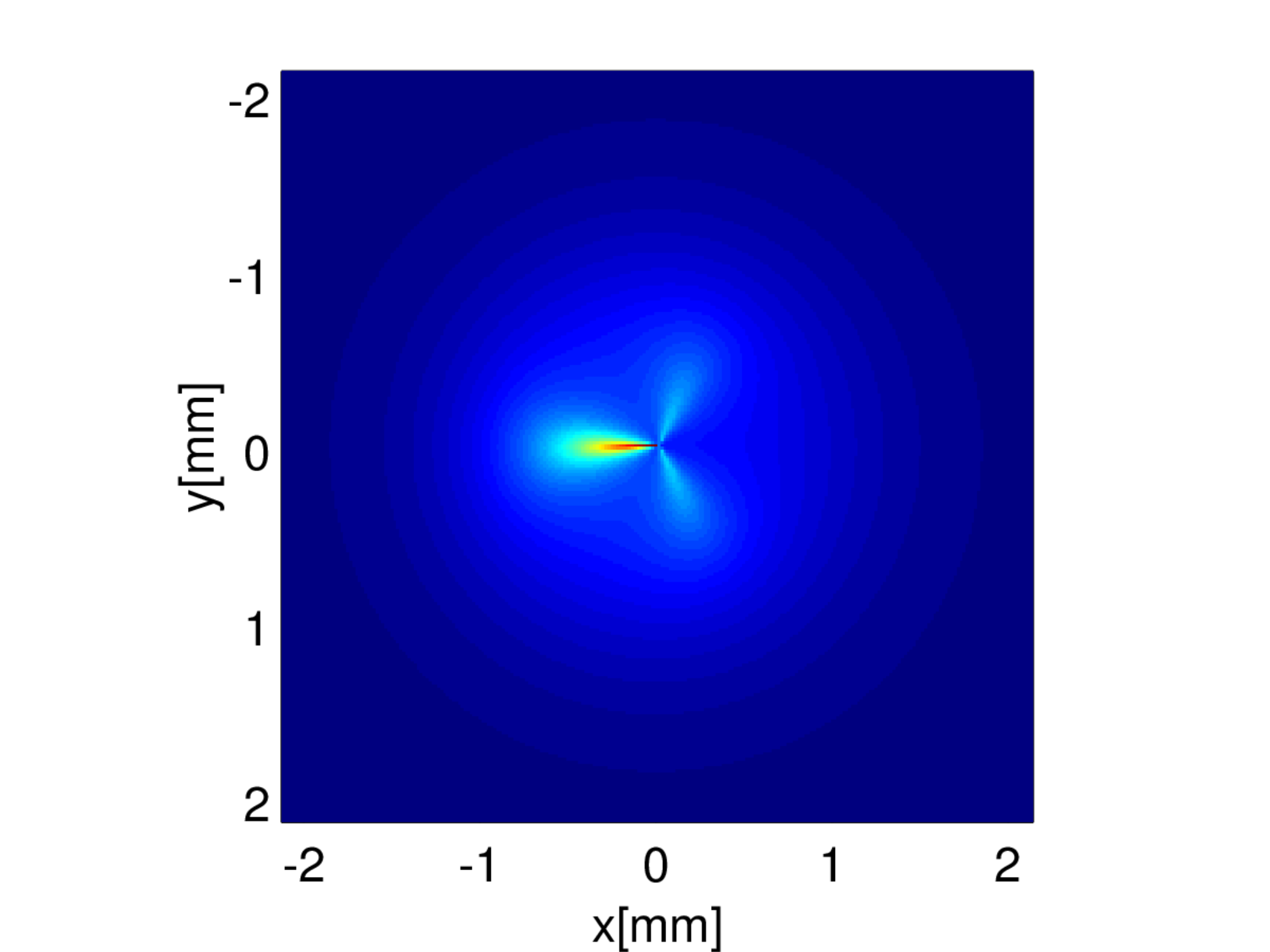}
\includegraphics[width=.31\textwidth]{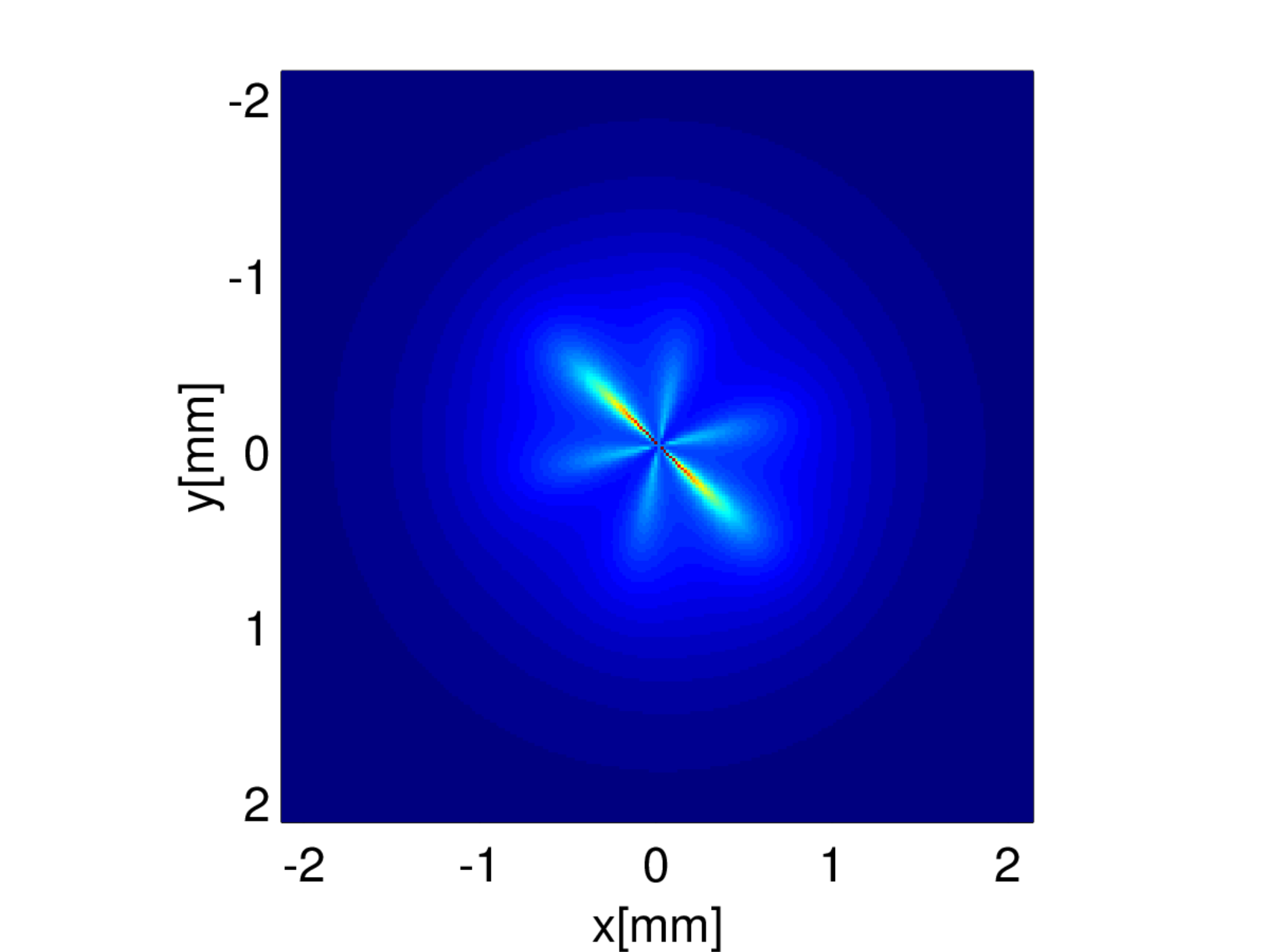}
\includegraphics[width=.31\textwidth]{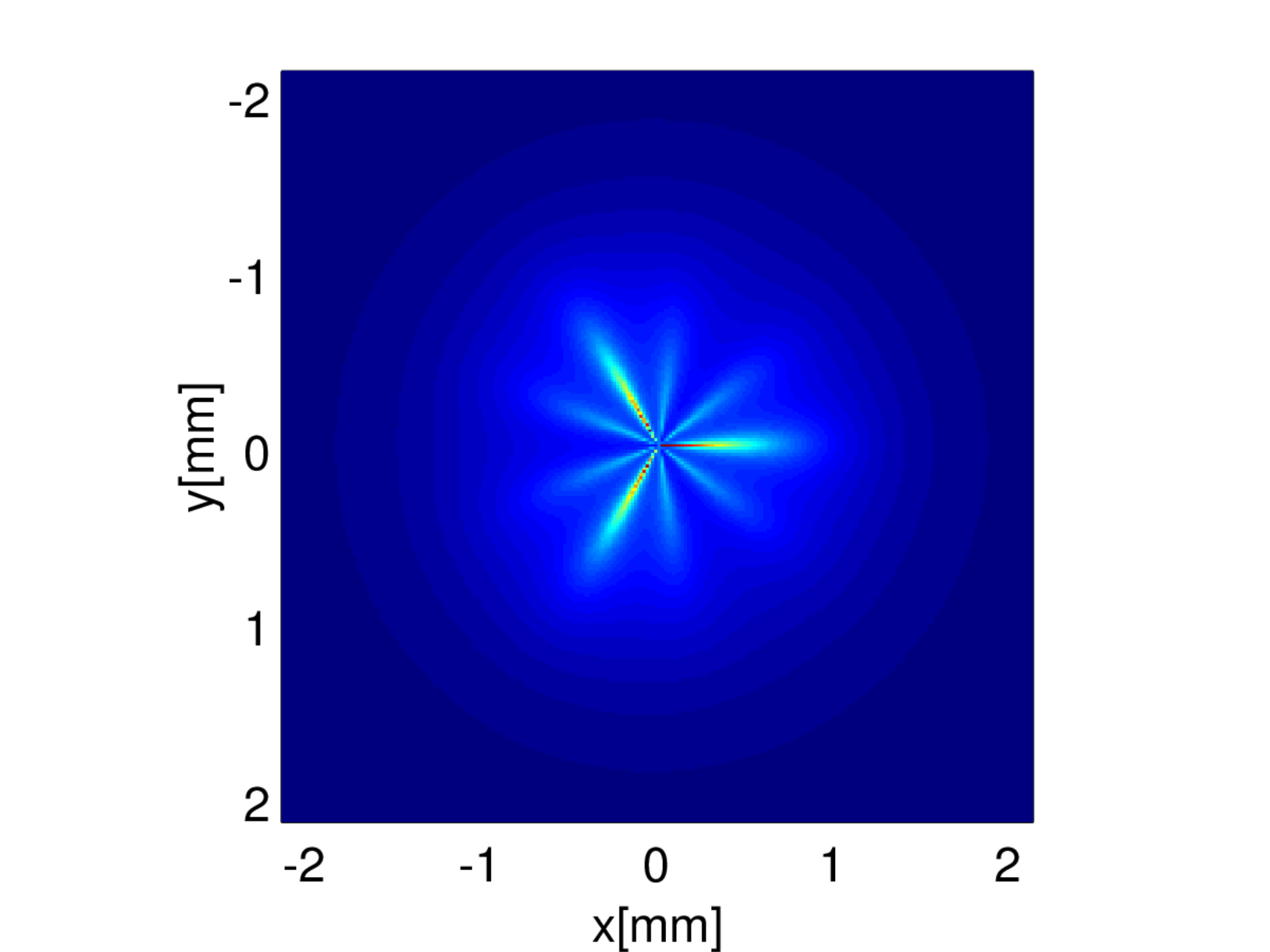}

(b)
\includegraphics[width=.31\textwidth]{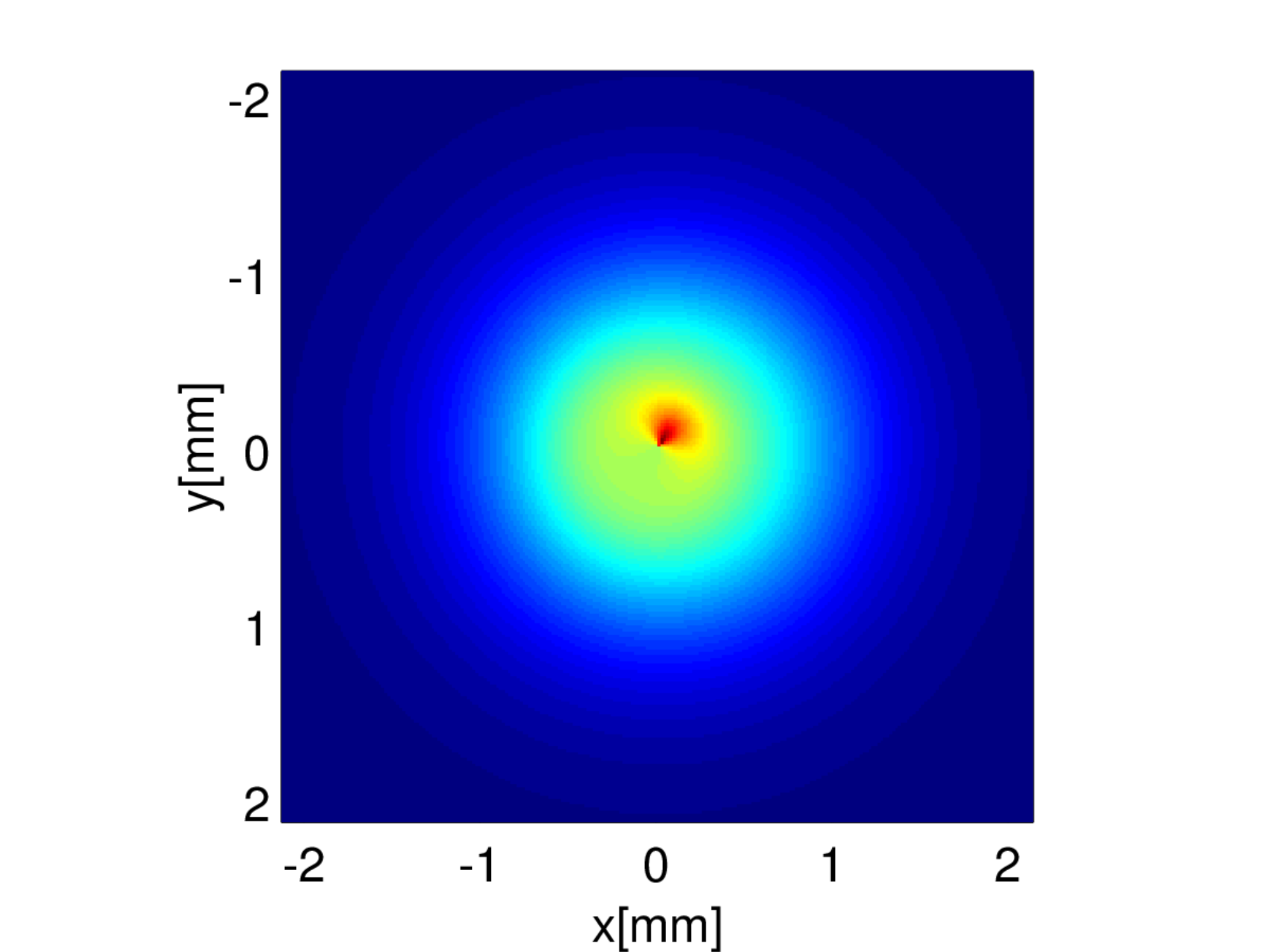}
\includegraphics[width=.31\textwidth]{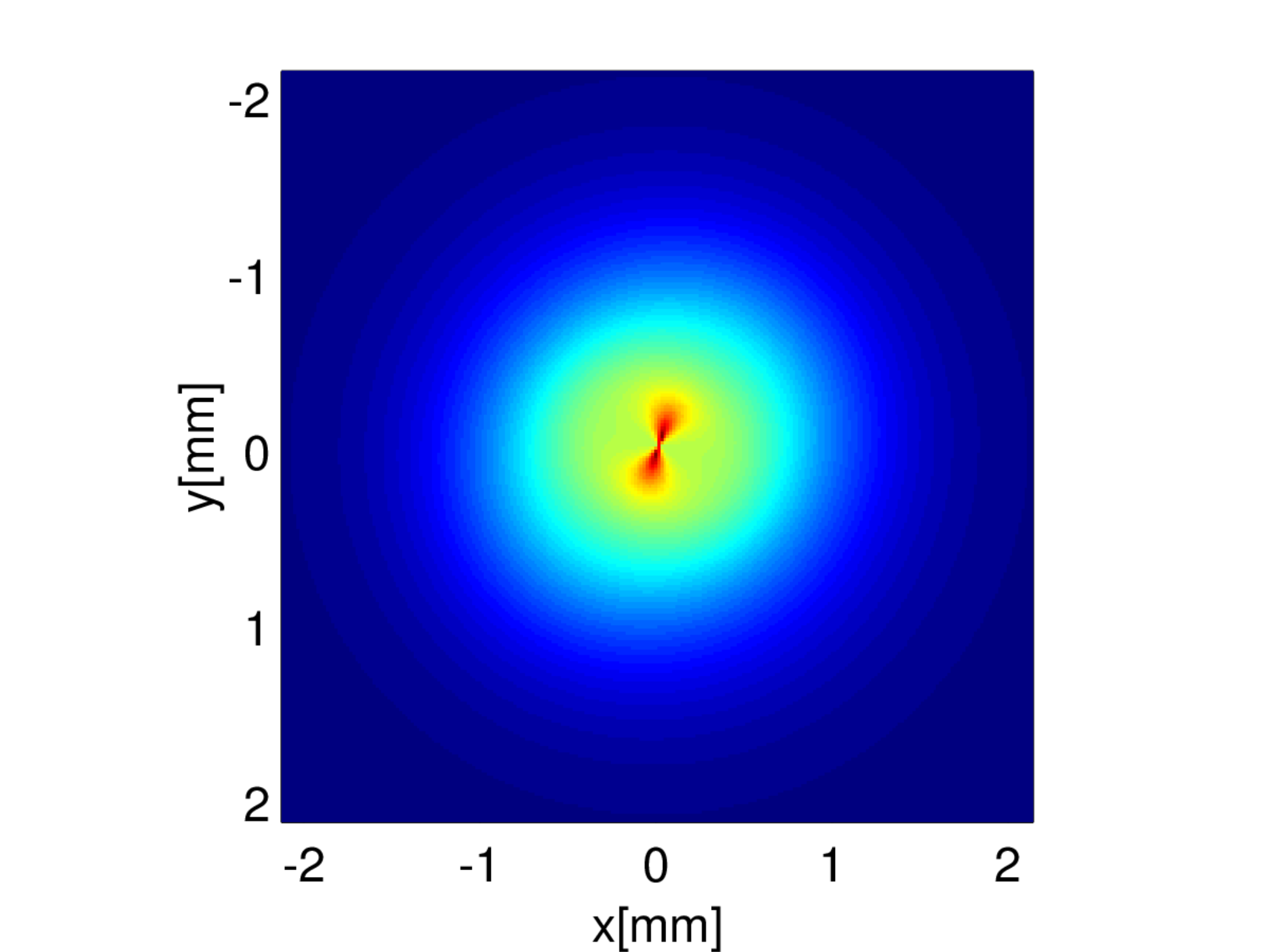}
\includegraphics[width=.31\textwidth]{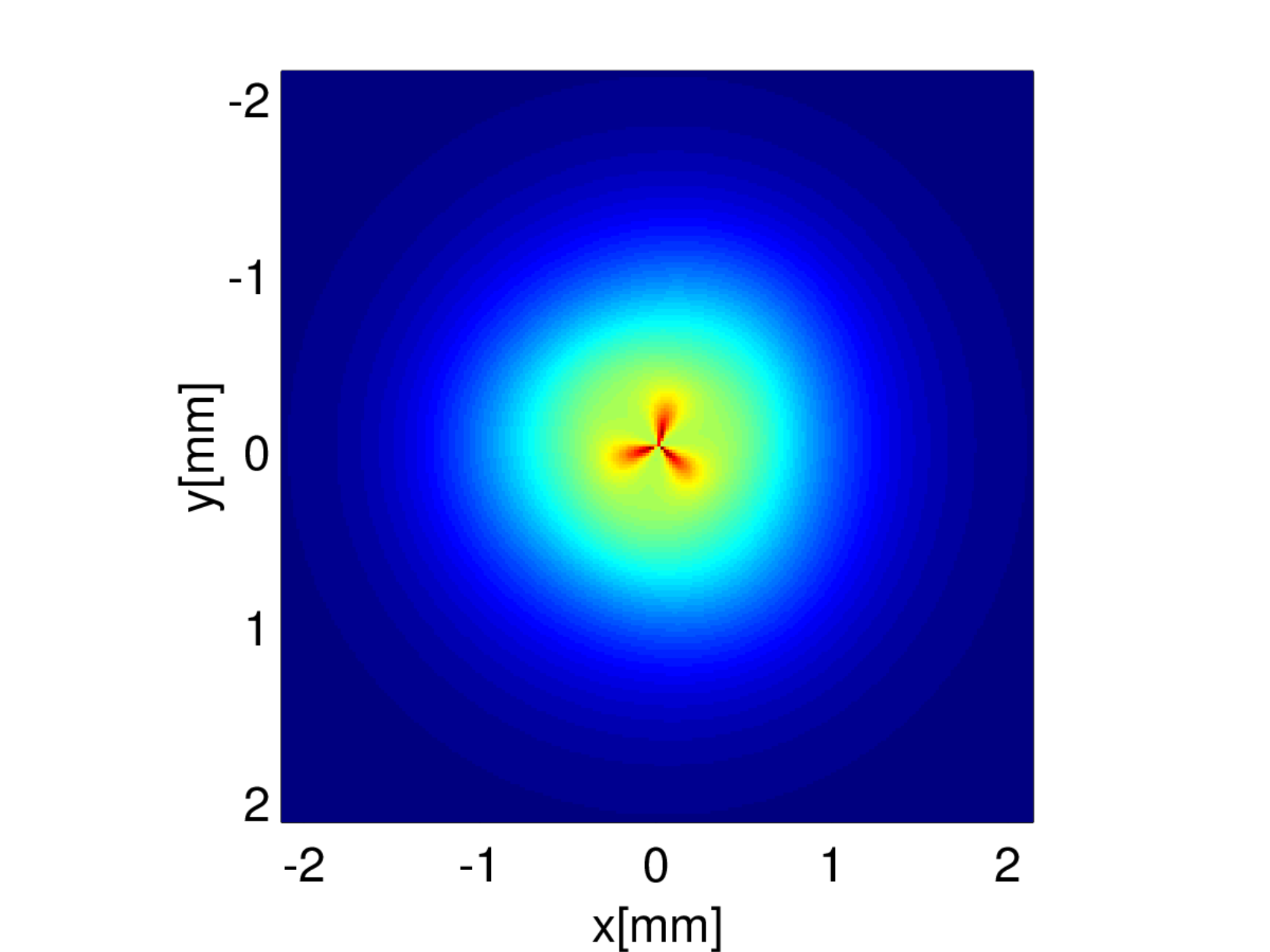}
\caption{The patterns for $Q=1,2,3$ vortex (left to right), in a dissipative (a) and lossless (b) left-handed metamaterial. All parameters are the same as in Fig.~\ref{pattvoc2} but the boundary condition at $z=0$ is now a vortex for
the electric field $E$ and a homogenous background for $H$. The symmetries and the whole qualitative picture are the same as before, confirming that the predictions of the paper do not require preparing a vortex in magnetic field at
the entry.}
\label{pattvoc3}
\end{figure}

Here we show that our results stay valid also when only the electric field has vortex patterns whereas the magnetic field starts analytic everywhere. As we discuss in the main text, this situation is experimentally more relevant than
the one assumed in most calculations in the paper (that both the electric and the magnetic field have a vortex as they enter the material). The electric field is typically a few orders of magnitude more intense than the magnetic field, as
seen in \cite{zarov03}. Therefore, one typically controls the electric field directly, imposing a given boundary condition at the front end of the material. Despite this fact, the magnetic field remains very important: the coupled
equations of motion (\ref{eome}-\ref{eomh}) require both $E$ and $H$ to be nonzero. Indeed, as explained in \cite{zarov03}, the left-handedness comes as a consequence of the hysteresis-type dependence of the magnetic permittivity on
$H$. So while it is crucial that $E$ and $H$ are both nonzero, it is also true that the results should remain valid for $\vert H\vert\ll\vert E\vert$, and for the boundary condition that only has a vortex in $E$ at the front of the
metamaterial, not for $H$. With such boundary conditions and the same parameter values as before, Fig.~\ref{pattvoc3} repeats the calculations of Fig.~\ref{pattvoc2}. Obviously, the symmetries remain the same and the similarity of
the results for the two cases is striking. Obviously, the $\vert E\vert^2$ map is insensitive to the details of the initial magnetic field pattern, as one expects from experiments and common wisdom in nonlinear optics.
We are thus content that the numerically simplifying assumption of identical $z=0$ boundary conditions for $E$ and $H$ does not put into question the findings of our work.

\section{The calculation of the self-energy diagrams}\label{appc}

We discuss here in some more detail the equations (\ref{loopcorrel}) from the main text. First we give the expressions for the couplings $g_{2,0,0},g_{0,2,0},g_{2,0,2},g_{0,2,2}$, which come from the expansion over the magnetic field $H$ of the nonlinear dependence $\mu(H)$ in (\ref{muloop}):
\bea
g_{2,0,0}=\frac{\alpha E_c^4\omega_0^2-(\omega-\imath\Gamma)\omega\alpha E_c^8}{H_0+\alpha E_c^4(\omega_0^2-(\omega-\imath\Gamma)\omega\alpha E_c^2)}\\
g_{0,2,0}=(k^2-\lambda^2)g_{2,0,0}\\
g_{2,0,2}=2\alpha E_c^2H_0\frac{\omega_0^2-(\omega-\imath\Gamma)\omega\alpha E_c^4}{\left(H_0+\alpha E_c^4(\omega_0^2-(\omega-\imath\Gamma)\omega\alpha E_c^2)\right)^2}\\
g_{0,2,2}=(k^2-\lambda^2)g_{2,0,2}.
\eea
For simplicity, we will treat the case when $\lambda=k$ and thus $g_{0,2,0}=g_{0,2,2}=0$. This simplifies the calculations substantially while it does not change the symmetry of the solution. It is possible to evaluate the diagram $\Sigma^{(1)}$ exactly in terms of sine and cosine integrals $\mathrm{Si}$ and $\mathrm{Ci}$. The angular integration is straightforward, the integration over $u$ results in four combinations of the trigonometric integrals, for the four terms in (\ref{prop}). Three of the four integrals are finite and therefore they just shift the mass term. The third term of the propagator is logarithmically divergent:
\be
\Sigma^{(1)}_3=\frac{4\pi\sin\pi Q}{Q\Gamma(Q/2)}e^{-3\imath\pi Q/2}\frac{1}{a^2}
\left(\gamma_E+\log\Lambda+(-1)^{Q}a\left(\cos\left(a\Lambda\right)\mathrm{Ci}\left(a\Lambda\right)+\sin\left(a\Lambda\right)\mathrm{Si}\left(a\Lambda\right)\right)\right).
\ee
To judge the effect of this term, we should extract the mass squared $r_m$ of the bare propagator, writing it out for small $u$:
\be
G(\mathbf{u}\to 0)=\frac{2\pi}{\Gamma(Q/2)}\frac{1}{u(u^2-a^2)}
\left(e^{\imath Q(\pi/2+\phi)}\left(\cos(a\Lambda-\pi Q)-\sin(a\Lambda)\right)+e^{-\imath Q(\pi/2+\phi)}\left(\cos(a\Lambda+\pi Q)-\cos(a\Lambda)\right)\right).
\ee
Since $G^{-1}(\mathbf{u}\to 0)\propto u=0$, the bare propagator is massless. The one-loop correction $\Sigma^{(1)}$ therefore gives a cutoff-dependent mass $r_M\sim\log\Lambda$, which could be absorbed in the overall normalization of the propagator. As we declared in the main text, the one-loop self-energy does not do much.

The crucial diagram $\Sigma^{(2)}$, the popular watermelon diagram, cannot be calculated exactly. It can be evaluated in the regime of small external momentum $\mathbf{u}$, i.e, when $u<u^\prime,u^{\prime\prime}$; more precisely, we can look at the regime when $u<u_0<u^\prime,u^{\prime\prime}$ for some (arbitrary) scale $u_0$ and expand in a series in $u/u_0$. Let us denote such entity by $\Sigma^{(2)}(\mathbf{u};u_0)$: it contains enough information for our purposes: we are interested mainly in angular integrations which determine the symmetry, and these can be done exactly as they separate from the integrations over the module $u$ in the small-$u$ limit. For $\mathbf{u}=0$ the watermelon diagram reads (with $\int\equiv\int_0^{2\pi}d\phi^\prime\int_0^{2\pi}d\phi^{\prime\prime}\int du^\prime\int du^{\prime\prime}$):
\bea
\nonumber\Sigma^{(2)}&\approx &\int\frac{G(\mathbf{u}^\prime)G(\mathbf{u}^{\prime\prime})}{v}
\left(e^{\imath Q(\pi/2+(\phi-\phi^\prime-\phi^{\prime\prime}))}(\cos(a\Lambda-\pi Q)-\sin(a\Lambda))+e^{-\imath Q(\pi/2+(\phi-\phi^\prime-\phi^{\prime\prime}))}(\cos(a\Lambda+\pi Q)-\cos(a\Lambda))\right)\\
v&\equiv&\sqrt{(u^\prime)^2+(u^{\prime\prime})^2-2u^\prime u^{\prime\prime}\cos(\phi-\phi^\prime-\phi^{\prime\prime})}.
\eea
One angular integration is performed by taking $\phi^\prime\mapsto\phi^\prime+\phi^{\prime\prime}$, which makes the $\phi^{\prime\prime}$ integral completely trivial, and the $\phi^\prime$ integral is evaluated in terms of the elliptic integrals $E,K$. The outcome is finite, hence it is observable (not only at the cutoff scale), and reads:
\bea
\nonumber\Sigma^{(2)}(0;u_0)&=&\left(\frac{2\pi)}{a\Gamma(Q/2)}\right)^3e^{3\imath Q/2}\cos(3Q\phi/2)^2\int du^\prime \int du^{\prime\prime}
\frac{((u^\prime)^2-(u^{\prime\prime})^2)(u^\prime+u^{\prime\prime})E\left(-\frac{4u^\prime u^{\prime\prime}}{(u^\prime+u^{\prime\prime})^2}\right)}{(u^\prime)^2(u^{\prime\prime})^2((u^\prime)^2-a^2)((u^{\prime\prime})^2-a^2)((u^\prime)^2-(u^{\prime\prime})^2)}=\\
&=&\frac{1}{4\pi}\left(\frac{2\pi}{a\Gamma(Q/2)}\right)^3e^{3\imath Q/2}\cos(3Q\phi/2)^2(a^{3/2}-1/\Lambda^{3/2})+O\left(1/\Lambda^2\right)
\eea
In particular, this means that a nontrivial mass term is acquired, of the order $a^{3/2}$. This mass is anisotropic, and the factor $\cos(3Q\pi/2)^2$ is all we need for the $3Q$-polygon. The leading correction in $u/u_0$ is in fact inessential for the symmetry, but it is important as it contains a nonzero imaginary part, introducing a finite lifetime for such patterns. It reads
\bea
\nonumber\Sigma^{(2)}(u;u_0)&=&\int\frac{G(\mathbf{u}^\prime)G(\mathbf{u}^{\prime\prime})}{w}
\left(e^{\imath Q(\pi/2+(\phi^\prime-\phi^{\prime\prime}))}(\cos(a\Lambda-\pi Q)-\sin(a\Lambda))+e^{-\imath Q(\pi/2+(\phi^\prime-\phi^{\prime\prime}))}(\cos(a\Lambda+\pi Q)-\cos(a\Lambda))\right)=\\
\nonumber &=&\frac{1}{4\pi}\left(\frac{2\pi}{a\Gamma(Q/2)}\right)^3e^{3\imath Q/2}\left(\frac{2\imath a^{3/2}}{\pi}\sin(3Q\phi/2)+\frac{2\Lambda^{3/2}}{\pi}\cos(3Q\phi/2)\right)\\
w&\equiv&\sqrt{(u^\prime)^2+(u^{\prime\prime})^2-2u^\prime u^{\prime\prime}\cos(\phi^\prime-\phi^{\prime\prime})-2u\left(u^\prime\cos(\phi-\phi^\prime)+u^{\prime\prime}\cos(\phi-\phi^{\prime\prime})\right)}
\eea
At leading order, this tedious expression behaves like $1/r^3$, falling off much quicker than the bare propagator (\ref{prop}), which goes as $1/\sqrt{r}$ (most obvious from the Bessel-function form of the real-space solution), suggesting that the shape of the vortex, which is mainly determined by long-distance behavior, is not much influenced by the finite-$u$ correction to $\Sigma^{(2)}$.

\end{document}